\documentclass{aa}

\usepackage{graphicx}
\usepackage{txfonts}

\newcommand{\Mpc}{$h^{-1}$\thinspace Mpc}
\newcommand{\hmpc}{$h$\thinspace Mpc$^{-1}$}
\newcommand{\etal}{{\rm et al.~}}

\newcommand{\be}{\begin{equation}}
\newcommand{\ee}{\end{equation}}

\def\apj{ApJ}
\def\apjl{ApJL}
\def\apjs{ApJS}

\edef\aap{A\&A}

\begin{document}   

\title{ The richest superclusters} 
\subtitle{I. Morphology} 

\author{ M. Einasto\inst{1} \and E. Saar\inst{1} 
\and L. J. Liivam\"agi\inst{1} \and J. Einasto\inst{1} 
\and E. Tago\inst{1} \and
V.J. Mart\'{\i}nez$^{2}$ \and J.-L. Starck$^{3}$ 
\and V. M\"uller\inst{4} \and P. Hein\"am\"aki\inst{5}  
\and P. Nurmi\inst{5} 
\and M. Gramann\inst{1}  
\and  G. H\"utsi\inst{1} 
}

\institute{Tartu Observatory, EE-61602 T\~oravere, Estonia
\and
Observatori Astron\`omic, Universitat de Val\`encia, Apartat
de Correus 22085, E-46071 Val\`encia, Spain 
\and 
CEA-Saclay, DAPNIA/SEDI-SAP, Service d'Astrophysique, F-91191 Gif
sur Yvette, France 
\and
Astrophysical Institute Potsdam, An der Sternwarte 16,
D-14482 Potsdam, Germany
\and
Turku University,
Tuorla Observatory, V\"ais\"al\"antie 20, Piikki\"o, Finland 
}

\date{ Received 2007; accepted} 

\authorrunning{M. Einasto et al.}

\titlerunning{2dF rich superclusters}

\offprints{M. Einasto }

\abstract 
{Superclusters are the largest systems in the Universe to give us information
  about the formation and evolution of structures in the very early Universe.
  Our present series of papers is devoted to the study of the morphology and
  internal structure of superclusters of galaxies.  }
{ We study the morphology of the richest superclusters
  from the catalogues of superclusters of galaxies in the 2dF Galaxy
  Redshift Survey and compare the morphology of real superclusters with model
  superclusters in the Millennium Simulation.  }
{We use Minkowski functionals and shapefinders to quantify the morphology of
  superclusters: their sizes, shapes, and clumpiness.  We generate empirical
  models of simple geometry to understand which morphologies correspond
  to the supercluster shapefinders.  }
{Rich superclusters have elongated, filamentary shapes with high-density
  clumps in their core regions.  The clumpiness of superclusters is determined
  using the fourth Minkowski functional $V_3$. In the $K_1$-$K_2$ shapefinder
  plane the morphology of superclusters is described by a curve which is
  characteristic to multi-branching filaments. We also find that the
  differences between the fourth Minkowski functional $V_3$ for the bright and
  faint galaxies in observed superclusters are larger than in simulated
  superclusters.  }
{Our results show how the Minkowski functionals and shapefinders describe the
  morphology of superclusters.  We see that the observed superclusters are
  more diverse than model superclusters. There are a number of
  differences between observed and model superclusters,
   especially in the distribution of bright and faint galaxies.  }


\maketitle

\section{Introduction}

Superclusters of galaxies with characteristic dimensions of up to 100~\Mpc\ 
\footnote{$h$ is the Hubble constant in units of 100~km~s$^{-1}$~Mpc$^{-1}$.} 
are the largest relatively isolated density enhancements in the Universe.
Superclusters give us information about the early evolution of the structure
in the Universe (Kofman et al. \cite{kofman87}) and about the properties of
the supercluster -- void network at present days.

Early relatively deep all-sky catalogues of superclusters of galaxies 
were complied by Zucca \etal (\cite{z93}) and Einasto \etal 
(\cite{e1994}, \cite{e1997}, \cite{e2001}) on the basis of Abell 
clusters of galaxies (Abell \cite{abell}, Abell et al. \cite{aco}). New 
deep redshift surveys of galaxies (the Las Campanas Redshift Survey, the 
Sloan Digital Sky Survey and the 2 degree Field Galaxy Redshift Survey) 
cover large regions of sky and allow to investigate the distribution of 
galaxies and the properties of galaxies up to rather large distances 
from us. These surveys have served as the basis for compiling catalogues of 
superclusters of galaxies (Einasto et al. (\cite{e03a}; \cite{e03b}); 
Basilakos (\cite{bas03}); and Erdogdu et al. (\cite{erd04}). 

On the basis of the 2dF Galaxy Redshift Survey, we recently compiled a new
catalogue of superclusters of galaxies -- Einasto et al. (\cite{e06a},
hereafter Paper I). In this study we also compiled a catalogue of
superclusters of the Millennium Simulation by Springel et al.
(\cite{springel05}), and used this catalogue to study possible selection
effects.  In Einasto et al.  (\cite{ e06b}, hereafter Paper II) we studied the
properties of superclusters.  There we characterized overall geometry of
superclusters by their sizes, the degree of asymmetry and compactness, and
compared those with similar parameters of simulated superclusters. In Einasto
et al. (\cite{e07c}, Paper III) we discussed properties of galaxies in
superclusters, compared the density distributions and the properties of galaxy
populations in rich and poor superclusters. We also compared the luminosity
and multiplicity functions of observed and simulated superclusters (Einasto et
al. \cite{e06c}).

Our studies showed that there exist several differences between rich and poor
superclusters: rich superclusters contain high density cores which are absent
in poor superclusters. Rich superclusters have a larger fraction of passive,
red, non-star-forming galaxies than poor superclusters.  Interestingly, we
found that the fraction of very luminous superclusters in observed catalogues
is larger than in simulated catalogues.

Therefore, among all superclusters, the richest superclusters deserve 
special attention. The richest relatively nearby superclusters are the 
Shapley Supercluster (Proust et al. \cite{proust06} and references 
therein) and the Horologium--Reticulum Supercluster (Rose et al. 
\cite{rose02}; Fleenor et al.  \cite{fleenor05}; Einasto et al. 
\cite{e03d}).  Very rich superclusters began to form 
earlier than other structures, they are sites of early star and 
galaxy formation (e.g. Mobasher et al. 
\cite{mob05}) and first places 
where systems of galaxies form 
(e.g. Ouchi et al. \cite{ouch}, Venemans et al. \cite{ven} and others). 
The supercluster environment affects the properties of groups and clusters 
located there (Plionis \cite{pl04}). The fraction of X-ray clusters in 
rich superclusters is larger than in poor superclusters (Einasto \etal 
\cite{e2001}), and the core regions of the richest superclusters may 
contain merging X-ray clusters (Rose et al. \cite{rose02}; Bardelli et 
al. \cite{bar00}).

One of the goals of the forthcoming Planck satellite is the study 
of the large scale structure  using 
the Sunyaev-Zeldovich (SZ) effect. 
Cross-correlation of SZ selected and 
optically selected superclusters
(and rich superclusters in particular) is part of the planned
scientific work which will be done with Planck data.

In the present  papers our primary goal is to quantify 
the morphology of individual richest 2dFGRS superclusters in detail
(this paper) and to study substructures and galaxy populations in these
superclusters (Einasto et al. \cite{e07e}, hereafter RII).  

In Paper II (Einasto et al. 2007b) 
we presented a thorough review about earlier studies about
the shapes and sizes of superclusters. One possibility to characterize
the shape of an object was suggested by Sahni et al. (\cite{sah98}),
who introduced shapefinders on the basis of Minkowski
functionals. These shapefinders have been used before to estimate the
filamentarity of superclusters (Sheth et al.  \cite{sheth03},
Basilakos \cite{bas03}).  In the present paper we shall use the
Minkowski functionals and shapefinders to quantify the morphology of
observed and simulated rich superclusters. In contrast to most earlier
studies we shall calculate the Minkowski functionals for the whole
range of threshold densities, starting with the lowest density used in
the supercluster search, up to the peak density in the supercluster
core.  We determine the clumpiness of superclusters using the fourth
Minkowski functional $V_3$ and quantify the overall shape of
superclusters by the $K_1$--$K_2$ shapefinder curves (the
morphological signature). We also compare the Minkowski functionals
of bright and faint galaxies. We generate a series of geometrical
models which serve us a prototypes of morphology to simulate the
morphological signature of observed superclusters.

The paper is composed as follows. In Section 2 we describe the galaxy
data, the supercluster catalogue and the data on the richest
superclusters. In Section 3 we describe the Minkowski functionals and
shapefinders used to study the morphology of superclusters, and
present the results on supercluster morphology.  In Section 4 
we discuss our results and give our conclusions.
In the Appendix we introduce geometrical models as prototypes of morphology
to study the shapefinders, and describe different kernels used
to calculate the density fields of superclusters.

\section{Data}

\subsection{Catalogues of superclusters and groups}

We have used the 2dFGRS final release (Colless \etal \cite{col01};
\cite{col03}) that contains 245,591 galaxies. This survey has allowed
the 2dFGRS Team and many other groups to estimate the fundamental
cosmological parameters and to study intrinsic properties of galaxies
in various cosmological environments; see Lahav (\cite{lahav04} and
\cite{lahav05}) for recent reviews. We used the data about galaxies
and groups of galaxies (Tago et al.  \cite{tago06}, hereafter T06) to
compile a catalogue of superclusters of galaxies from the 2dF survey
(Paper I).  The 2dF sample becomes very diluted at large distances,
thus we restrict our sample by a redshift limit $z=0.2$; we apply a
lower limit $z \geq 0.009$ to avoid confusion with unclassified
objects and stars. When calculating (comoving) distances we use a flat
cosmological model with the 
standard parameters: matter density $\Omega_m =
0.3$, dark energy density $\Omega_{\Lambda} = 0.7$ (both in units of
the critical cosmological density).

Galaxies were included in the 2dFGRS, if their corrected apparent
magnitude ${\rm b_j}$ lied in the interval from $m_1 = 13.5$ to $m_2 =
19.45$. The faint limit actually fluctuates from field to field; these
fluctuations have been taken into account in the calculation of
weights assigned to galaxies. These weights were used to correct the
luminosities of galaxies. In the calculation of weights we used for
every galaxy the individual values of the faint end magnitudes of the
observational window, $m_2$. 
We also used a correction for the incompleteness factor $c = \gamma (1 -
\exp(m - \mu))$, where $\gamma = 0.99$, $m$ is the observed magnitude
of the galaxy, and the parameter $\mu$ varies from field to field (see
eq. 5 of Colless et al. \cite{col01}). The weight of the galaxy is
proportional to the inverse of the incompleteness factor.  To
calculate weights, we assumed that galaxy luminosities are distributed
according to the Schechter (\cite{S76}) luminosity function.
The weights are proportional to the ratio of the expected total luminosity
to the luminosity in the observational window of the survey at the
distance of the galaxy.

We used the weighted luminosities of galaxies to calculate the
luminosity density field 
on a grid with cell size of 1~\Mpc\ and smoothed with an
Epanechnikov kernel of radius 8~\Mpc; this density field was used to
find superclusters of galaxies. We defined superclusters as connected
non-percolating systems with densities above a certain threshold
density; the actual threshold density used was 4.6 in units of the
mean luminosity density. A detailed description of the supercluster
finding algorithm can be found in Paper I.

Later we shall use the data on the luminosities of galaxies to divide galaxies
by their luminosity into the populations of bright and faint galaxies. We
wanted to use an absolute magnitude limit close to the break luminosity
$M^\star$ in the Schechter luminosity function. According to the estimates of
the luminosity function, the value of $M^\star$ is different for different
galaxy populations (Madgwick et al.  \cite{ma03a}; de Propris et al.
\cite{depr03}; Croton et al.  \cite{cr05}); having values from $-19.0$ to
$-20.9$\footnote{all absolute magnitudes have been calculated for $h = 1$.}.
Therefore we used a bright/faint galaxy limit $M_{bj} = -20.0$ as a compromise
between the different values (see also Paper III).

For comparison we used simulated galaxy samples of the Millennium
Simulation by Springel et al. (\cite{springel05}). This simulation was
made using a very large number of dark matter particles 
($2130^3$) in
a periodic box of the size of 500~\Mpc, and adopting standard values
of cosmological parameters.  
For identifying superclusters in simulations, we adopted 
the same selection window as in the case of observed
superclusters (Paper I).  Using semi-analytic methods, catalogues of
simulated galaxies were calculated by Croton et al.  (\cite{cr06}). 
The simulated galaxy catalogue contains almost 9 million objects, for
which positions and velocities are given, as well as absolute
magnitudes in the Sloan Photometric system ({\tt u,g,r,i,z}).  The
limiting absolute magnitude of the catalogue is $-17.4$ in the {\tt r}
band.

The catalogues of observed groups and isolated galaxies
can be found at \texttt{http://www.aai.ee/$\sim$maret/2dfgr.html},
the catalogues of observed and model superclusters -- at
\texttt{http://www.aai.ee/$\sim$maret/2dfscl.html}.

\subsection{The richest superclusters in real and simulated catalogues}

{\scriptsize
\begin{table*}[ht]
\caption{Data on rich superclusters }
\tiny
\begin{tabular}{rrrrrlrrrrrcc} 
\hline 
 ID & R.A. & Dec & $D$ & $N_{gal}$ &$M_{lim}$ & $N_{vol}$ & $N_{cl}$
 &$N_{gr}$ &$N_{ACO}$&$N_X$ & $\delta_m$ &   $L_{tot}$ \\
     & deg & deg & \Mpc  &       &   &&  &  &     &   &              \\
\hline 
SCL88 (20)  & 155.11 & -2.54& 184.5  &   556  & -17.50& 484 & 2 &     7 &  1(7) &  2    & 5.7 &   0.319E+13 \\
SCL126 (152)& 194.71 & -1.74& 251.2  &  3591  & -19.25&1308 &18 &  40,2 &  9 &  5       & 7.7 &  0.378E+14 \\
SCL10 (5)   &   1.85& -28.06& 177.4  &   952  & -17.50& 757 & 5 &   5   & 1 (19) &  5   & 6.2 &   0.482E+13 \\
SCL9 (34)   &   9.85& -28.94& 326.3  &  3175  & -19.50&1176 &24 &  26,9 &  12 (25) &  6 & 8.1 &   0.497E+14\\
\\                      
M1 (195)    &       & &   &  5437 & -19.25&1589 &   9 &   &   &  &   8.2 &  0.204E+14\\
M2 (1089)   &       & &   &  5047 & -19.5 &4048 &   25 &   &   &  &  7.4 &  0.489E+14\\
M3 (1386)   &       & &   &  2016 & -19.5 &2007 &   17 &   &   &  &  7.2  &  0.638E+14\\
M4 (207)    &       & &   &  3645 & -19.5 &1794 &   9 &   &   &  &   7.8 &  0.581E+14\\
\label{tab:1}     
\end{tabular} 

Note: 
Identity ID after Einasto et al. (2001) with the name of Paper I in
parenthesis; with sky coordinates and distance $D$ for our cosmology; 
the galaxy number $N_{gal}$ for the whole superclusters,
and magnitude limit $M_{lim}$ and the galaxy number $N_{vol}$ for 
volume limited superclusters, $N_{cl}$ and $N_{gr}$ are density field 
cluster and group numbers according to paper I,  
$N_{ACO}$ and $N_X$ gives the number of Abell and X-ray clusters,
respectively, in that part of the supercluster covered by 2dF survey; 
the number inside parenthesis is the total number of Abell clusters in this
supercluster by Einasto et al. (2001) list;  
$\delta_m$ -- the mean values of the luminosity density field 
in superclusters, in units of mean density;
$L_{tot}$ -- supercluster total luminosity in Solar units.
\end{table*}            
}

For the present analysis we select from both our catalogues four rich 
superclusters. From the 2dF superclusters we chose two superclusters from 
the Northern and two from the Southern Sky. Two of them are the richest 
superclusters in our catalogue: the supercluster SCL126 in the Northern 
Sky, and the supercluster SCL9 (the Sculptor supercluster) in the 
Southern Sky, according to the catalogue by Einasto et al. 
(\cite{e2001}, hereafter E01). The others are two relatively nearby 
superclusters -- SCL88 (Sextans) in the Northern Sky, and SCL10 (the 
Pisces-Cetus supercluster) in the Southern Sky. All these superclusters 
are only partly covered by the 2dF Survey region (Einasto et al.  2001, 
2003).  There are several richer superclusters in the 2dFGRS 
supercluster catalogue, but they are far away, 
and due to the magnitude limit of the survey, we would only get
a small number of galaxies in volume limited supercluster galaxy samples. 
As we wanted to study the 
galaxy content of the superclusters in an accompanying paper, we had to 
select rich, but also relatively nearby superclusters.

A description of these superclusters is given in
Table~\ref{tab:1}.  
There we provide the coordinates and
distances of superclusters, the numbers of galaxies, groups and Abell
and X-ray clusters in the superclusters, the mean values of the
luminosity density field in the superclusters and their total
luminosities (from Paper II).  In our analysis we use volume-limited
samples of galaxies from these superclusters. The luminosity limits
for these samples for each supercluster are also given in
Table~\ref{tab:1}.

\begin{figure*}[ht]
\centering
\resizebox{0.22\textwidth}{!}{\includegraphics*{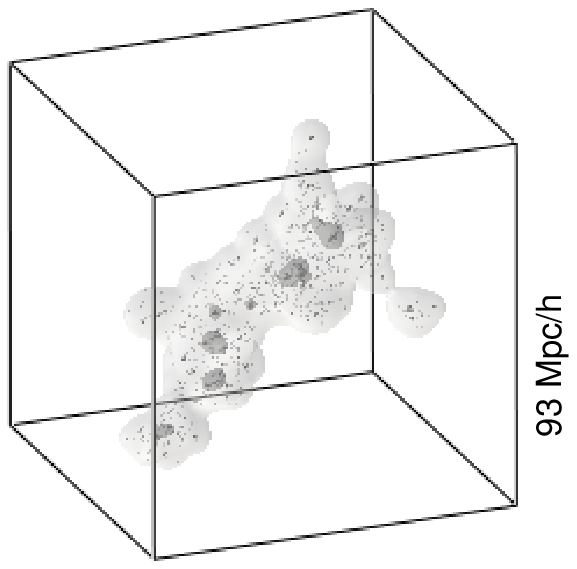}}
\resizebox{0.22\textwidth}{!}{\includegraphics*{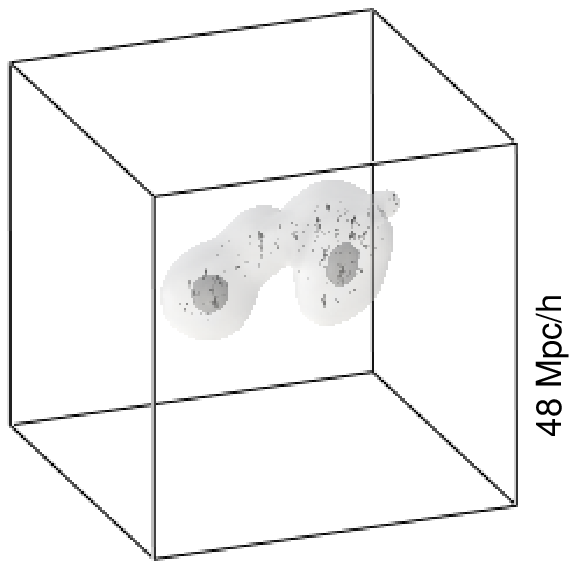}}
\resizebox{0.22\textwidth}{!}{\includegraphics*{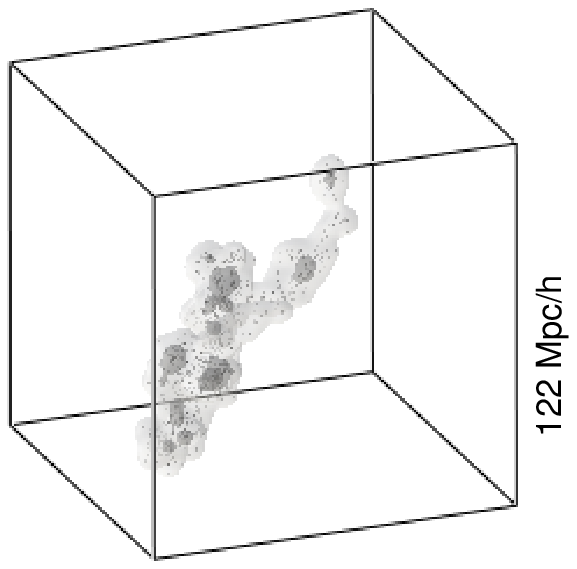}}
\resizebox{0.22\textwidth}{!}{\includegraphics*{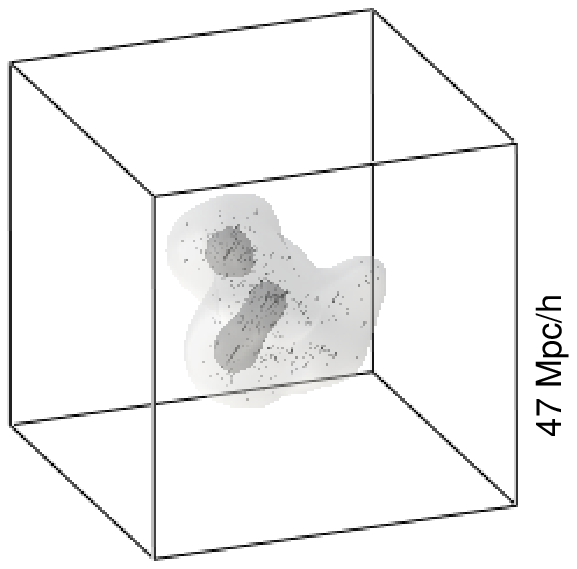}}
\hspace*{2mm}\\
\resizebox{0.22\textwidth}{!}{\includegraphics*{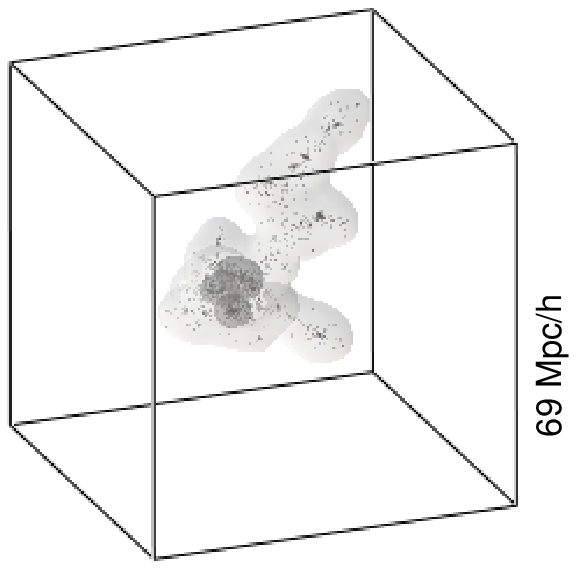}}
\resizebox{0.22\textwidth}{!}{\includegraphics*{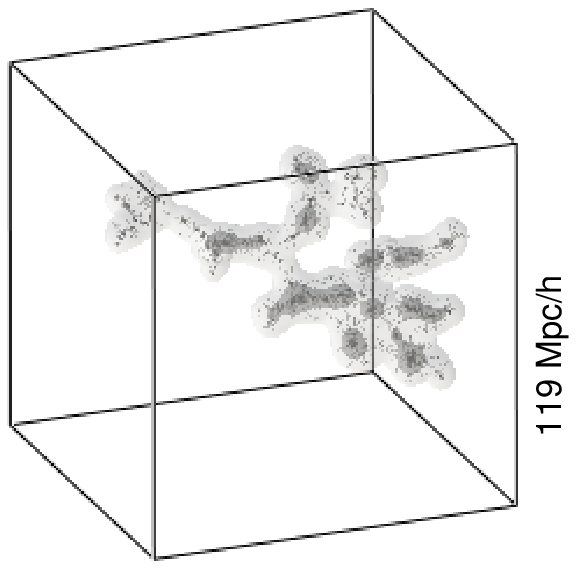}}
\resizebox{0.21\textwidth}{!}{\includegraphics*{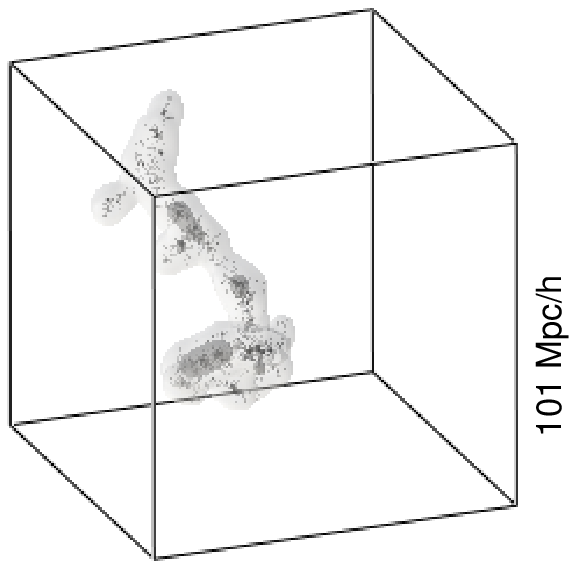}}
\resizebox{0.22\textwidth}{!}{\includegraphics*{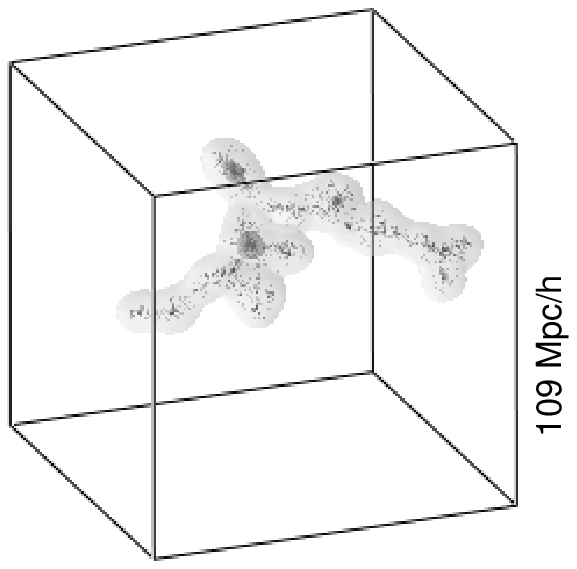}}
\caption{The density field view of the superclusters. We show with darker 
colours the central (core) region which approximately corresponds 
to the region for which the Minkowski functional
$V_{3}$ has a maximum. Galaxies are shown with
small dots. 
Upper panels from left to right: SCL126, SCL88, SCL9,  SCL10.
Lower panels from left to right: M1, M2, M3, M4. 
}
\label{fig:xydens}
\end{figure*}

The most prominent Abell  supercluster in the Northern 2dF survey is the 
supercluster SCL126 (in E01, N152 in Paper I) in the direction of the Virgo 
constellation. This supercluster has been also called the Sloan Great Wall 
(Vogeley et al. \cite{vogeley04}, Gott et al. \cite{gott05}, Nichol et al. 
\cite{nichol06}). 

Another rich supercluster in the Northern Sky is the Sextans supercluster, 
SCL88 (in E01; N20 in paper I). Only a small part of this supercluster
is located inside the 2dF survey volume, including one of seven Abell clusters
from this supercluster.

The richest  supercluster  in the Southern Sky is  the Sculptor 
supercluster (SCL9 in E01; S34). This supercluster contains also several 
X-ray clusters. This supercluster contains the largest number of Abell 
clusters in our supercluster sample, 25. However, only 12 of then are 
located in the region covered by the 2dF redshift survey.  

Another nearby prominent supercluster in the Southern sky is the
Pisces-Cetus supercluster (SCL10 in E01, 
S5 in paper I) which contains the
rich X-ray cluster, Abell 2734. Only one of 19 Abell clusters from
this supercluster is located inside the 2dF survey boundaries. This
supercluster was recently described as a rich filament of Abell
clusters by Porter and Raychaudhury (\cite{pr05}).

From the Millennium simulation, we use the data on three richest 
superclusters. The  supercluster M3 is 9th richest in the catalogue  by 
the number of galaxies, but the reason to include this system in our 
analysis is that this supercluster is the second richest by the number 
of density-field (DF) clusters in it. Density-field clusters (Paper II) 
are local maxima of the luminosity density, and a counterpart to  
real galaxy clusters.

For comparison, we shall use the data on the best-known nearby supercluster,
the Local Supercluster, 
denoted as V20 due to the chosen distance limit. 
The Local Supercluster represents a typical poor
supercluster of a rather small size, with one rich galaxy cluster, the Virgo
cluster, in the centre, surrounded by filaments of galaxies and poor groups.
The Local Group is located near the edge of the supercluster. The total
luminosity of the Local Supercluster is $L = 3 \times 10^{12} h^{-2}
L_{\sun}$, and its mass is $M = 1 \times 10^{15} h^{-1} M_{\sun}$. Most
superclusters in our catalogue of the 2dFGRS superclusters are of the Local
Supercluster type (Paper I). The data on the Local Supercluster galaxies are
taken from ZCAT (\texttt{http://cfa- www.harvard.edu/$\sim$huchra/zcat/}).  In
total we have in this supercluster 328 galaxies in a volume-limited sample ($M
\leq - 18.0$), with the maximum distance of 20~\Mpc.

The distribution of galaxies in regions of different density in real
and simulated superclusters is shown in Fig. \ref{fig:xydens}.
We see at a first glance already how filamentary all the
superclusters are.  The presence of several concentration centres, as
well as many high-density knots is also clearly seen.

\section{Morphology of superclusters} 
\subsection{Morphological descriptors} 

We characterize superclusters by their outer (isodensity) surface, 
and its enclosed volume. When increasing the isodensity level over the
threshold overdensity $\delta=4.6$ (sect. 2.1), we move into the
central parts of the supercluster.  The morphology and topology of the
isodensity contours is (in the sense of global geometry) completely
characterized by four Minkowski functionals.

For a given surface the four Minkowski functionals are, respectively:
\begin{enumerate}
\item the first Minkowski functional $V_{0}$ is the enclosed volume V, 
\item the second Minkowski functional
 $V_{1}$ is proportional to the area of the surface 
$S$, namely,
    \begin{equation} 
        V_1 = {1\over6} S;
    \end{equation}
\item the third Minkowski functional
 $V_{2}$ is proportional to the integrated mean curvature C, 
\begin{equation} 
        V_2 = \frac1{3\pi} C, \quad 
        C=\frac12\int_S\left(\frac1{R_1}+\frac1{R_2}\right)\,dS,
\end{equation}
where $R_1$ and $R_2$ are the two local principal radii of curvature.
\item the fourth Minkowski functional
 $V_{3}$ is proportional to the integrated Gaussian 
curvature (or Euler characteristic) $\chi$, 
  \begin{equation}
  V_3=\frac12\chi,\quad 
  \chi = \frac1{2\pi}\int_S\left(\frac1{R_1R_2}\right)dS.
  \end{equation}
\end{enumerate}

The Euler characteristic 
is simply related to the genus, $G$
\begin{eqnarray}
\label{genus}
   G&=&1-V_3.  
  \end{eqnarray}
The fourth Minkowski functional  gives us the number of isolated 
clumps (or voids) in the sample (Martinez et al. 2005; 
Saar et al. \cite{saar06}). One should beware of extra factors of 2 that
are sometimes seen in formulae like (\ref{genus}); this error has
crept into these two papers, too. With conventional normalization
there should be no extra factors in (\ref{genus}).

To characterize the shape of an object Sahni et al. (\cite{sah98}) and
Shandarin et al.  (\cite{sss04}) introduced shapefinders, a set of
combinations of Minkowski functionals: $H_1=3V/S$ (thickness),
$H_2=S/C$ (breadth), and $H_3=C/4\pi$ (length). These quantities have
dimensions of length and are normalized to give $H_i=R$ for a sphere
of radius $R$.  For a convex surface, the shapefinders $H_i$ follow
the inequalities $H_1\leq H_2\leq H_3$.  Prolate ellipsoids (pancakes)
are characterized by $H_1 << H_2 \approx H_3$, while oblate ellipsoids
(filaments) are described by $H_1 \approx H_2 << H_3$.
  
Additionally, Sahni et al. (\cite{sah98}) defined  two dimensionless
shapefinders $K_1$ (planarity) and $K_2$ (filamentarity): 
$K_1 = (H_2 - H_1)/(H_2 + H_1)$ and $K_2 = (H_3 -
H_2)/(H_3 + H_2)$.

Then, after Sahni et al. (\cite{sah98}),
the following shapes can be distinguished:
\begin{enumerate}
\item spheres with $H_1 = H_2 = H_3$, i.e. $K_1 = K_2 = 0$;
\item ideal filaments with $K_1 \approx 0$, $K_2 \approx 1$;
\item real filaments with $K_1 < K_2$;
\item ideal pancakes with $K_1 \approx 1$, $K_2 \approx 0$;
\item pancakes with $K_1 > K_2$;
\item ideal ribbons with  $K_1 \approx K_2 \approx 1$;
\item ribbons with $K_1 / K_2 \approx 1$.
\end{enumerate}

In the $(K_1,K_2)$-plane filaments are located near the $K_2$-axis,
pancakes near the $K_1$-axis, and ribbons along the diagonal, connecting 
the spheres at the origin with the ideal ribbon at $(1,1)$.

\subsection{Supercluster morphology}

\begin{figure*}[ht]
\centering
\resizebox{0.45\textwidth}{!}{\includegraphics*{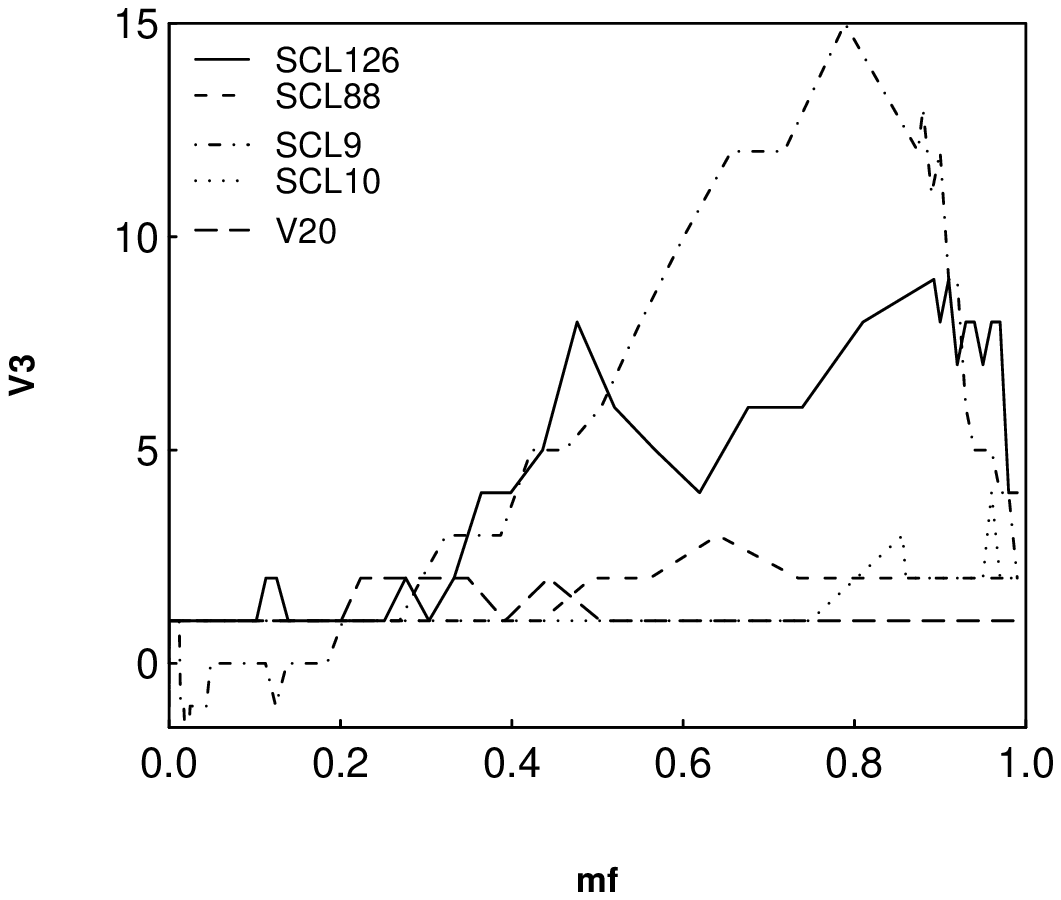}}
\resizebox{0.45\textwidth}{!}{\includegraphics*{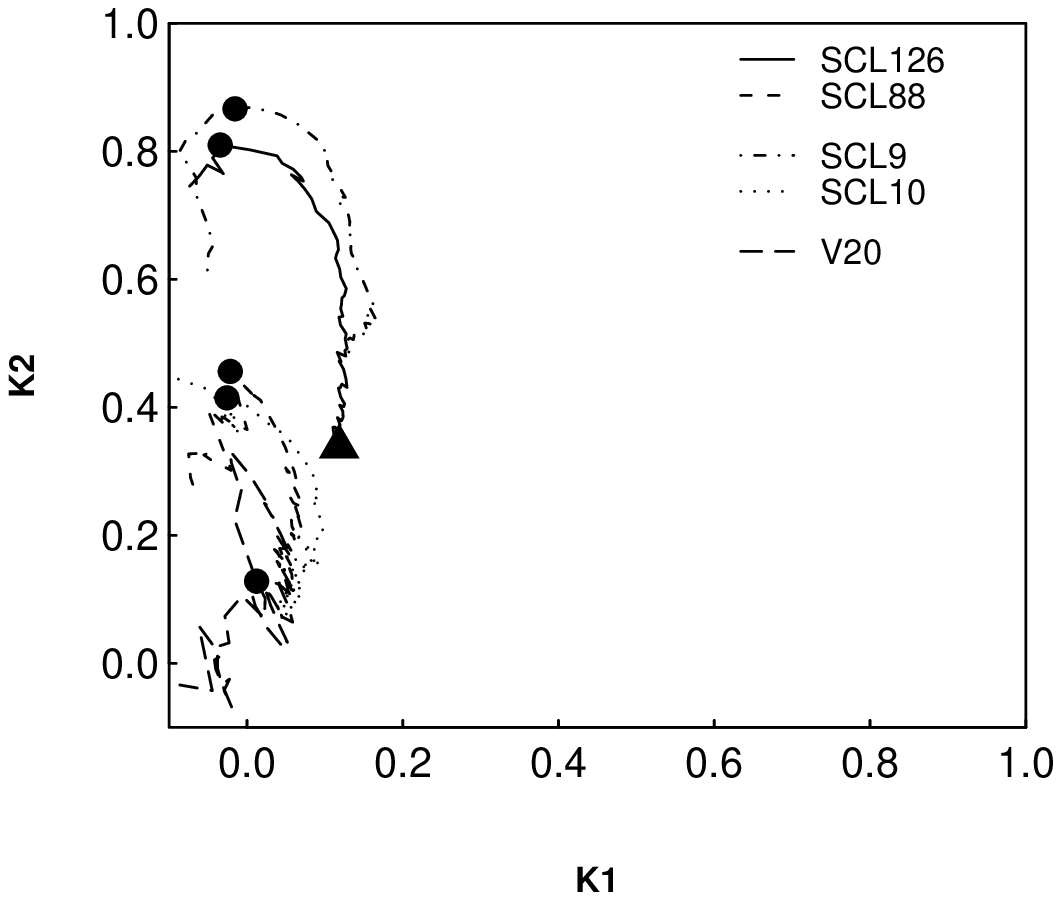}}
\hspace*{2mm}\\
\resizebox{0.45\textwidth}{!}{\includegraphics*{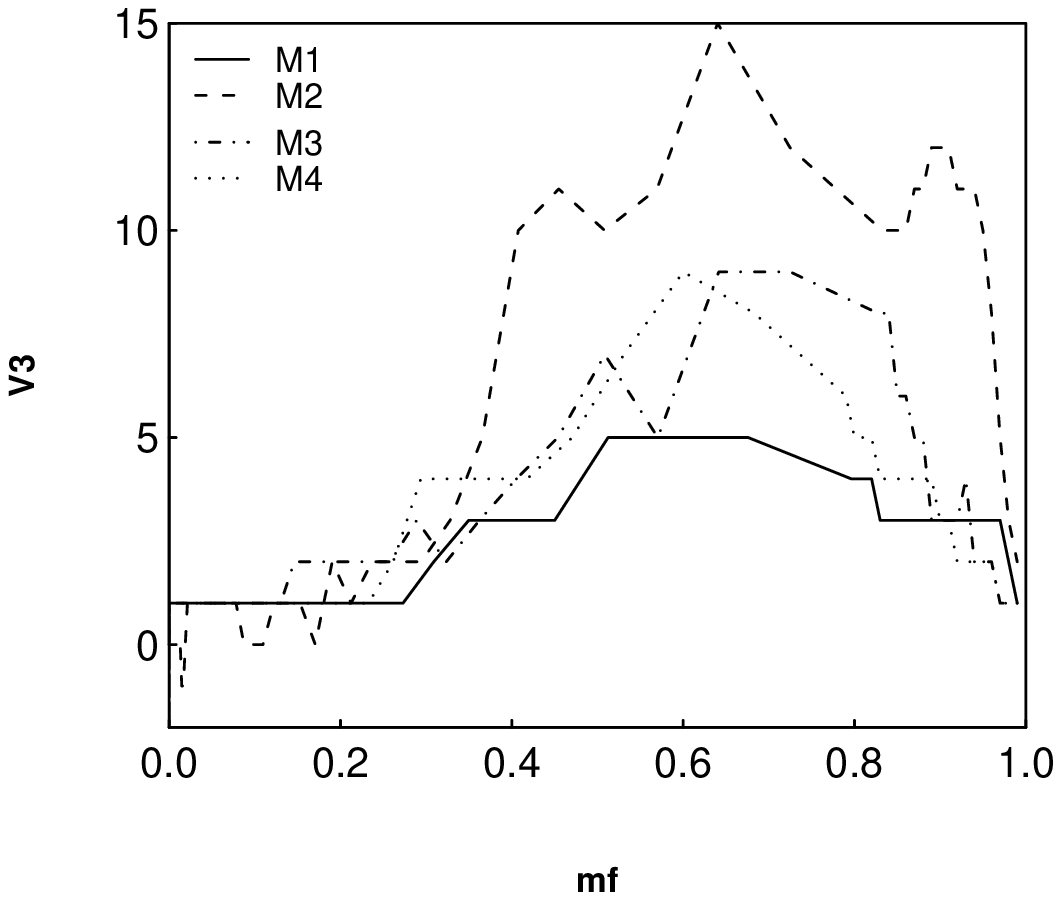}} 
\resizebox{0.45\textwidth}{!}{\includegraphics*{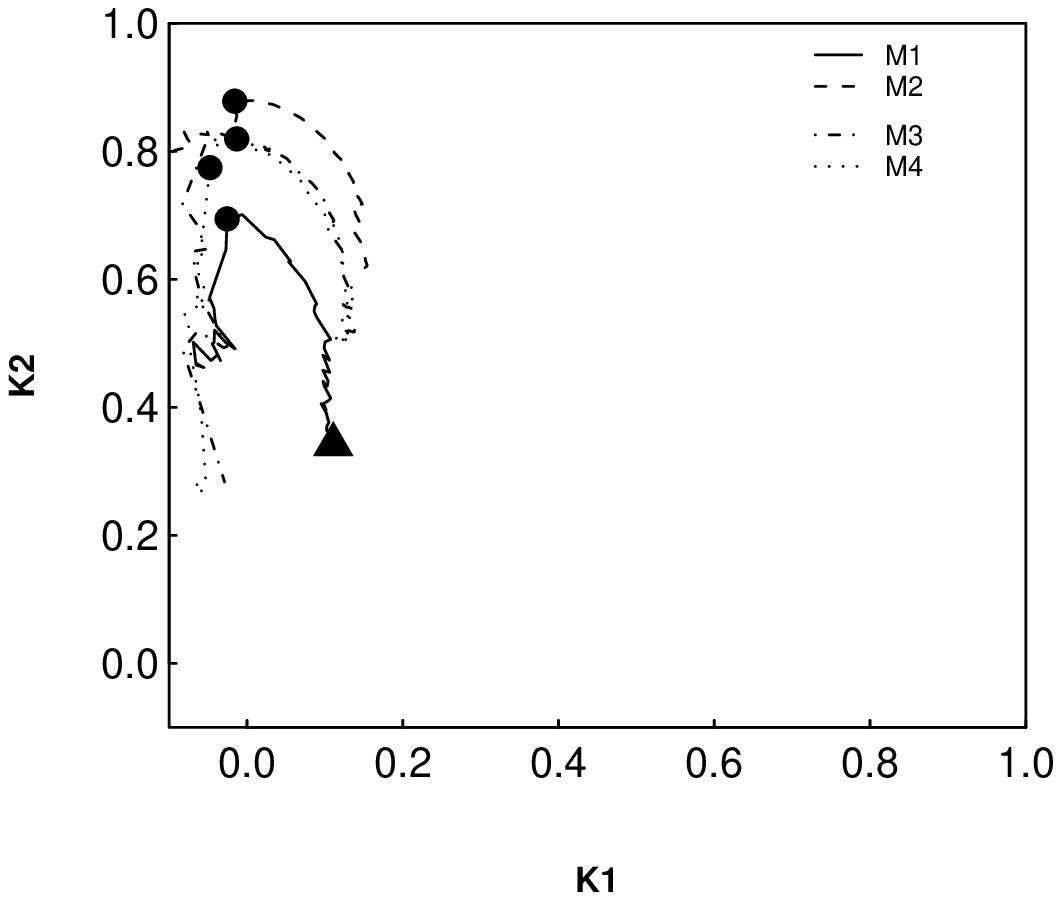}}
\caption{The Minkowski functional $V_{3}$ (the Euler characteristic)
  (left panels) and the shapefinders $K_1$ (planarity) and $K_2$
  (filamentarity) (right panels) for the observed (upper panels) and
  simulated (lower panels) superclusters.  In the right panels we
  indicate for SCL126 and M1 with triangles the values of $K_1,~K_2$,
  where the mass fraction $m_f = 0.0$ (the whole supercluster), and
  for all superclusters 
  with filled circles the values of $K_1,~K_2$, which correspond to
  the $m_f$, at which $V_3$ has a maximum.  }
\label{fig:mf4k1k2}
\end{figure*}

\begin{figure*}[ht]
\centering
\resizebox{0.28\textwidth}{!}{\includegraphics*{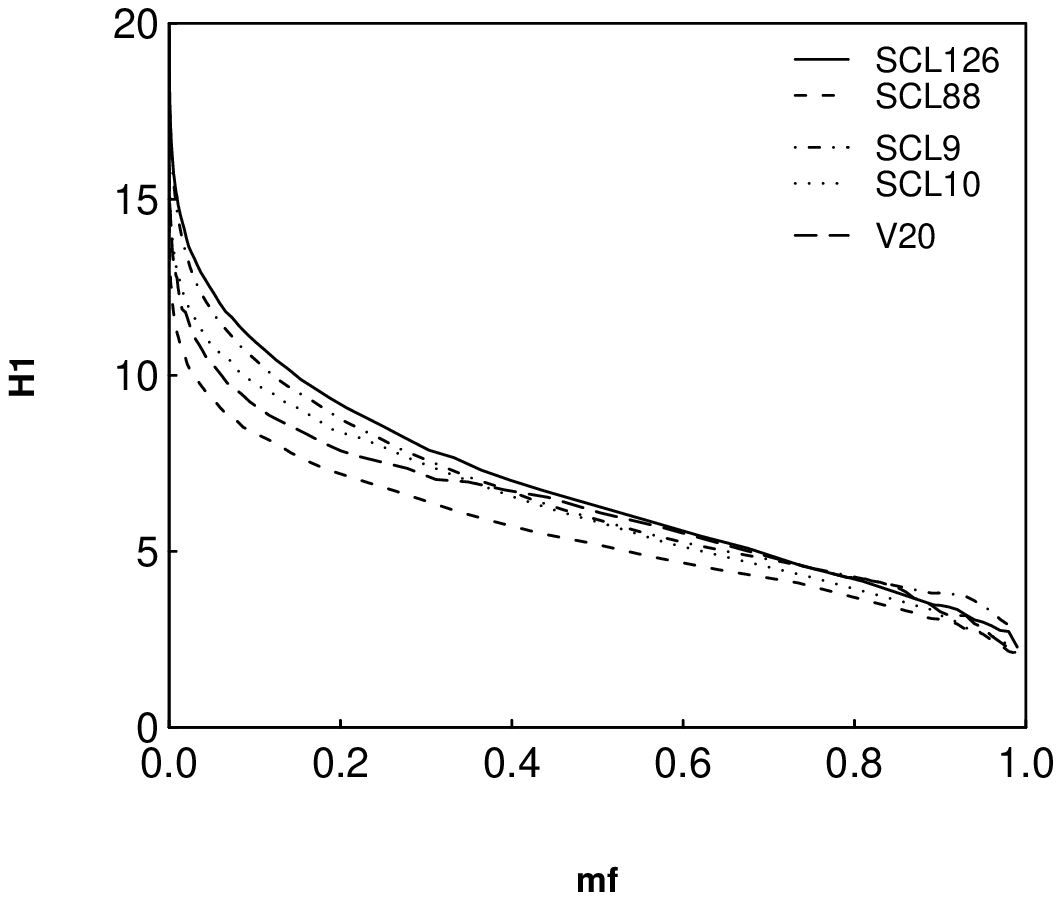}}
\resizebox{0.28\textwidth}{!}{\includegraphics*{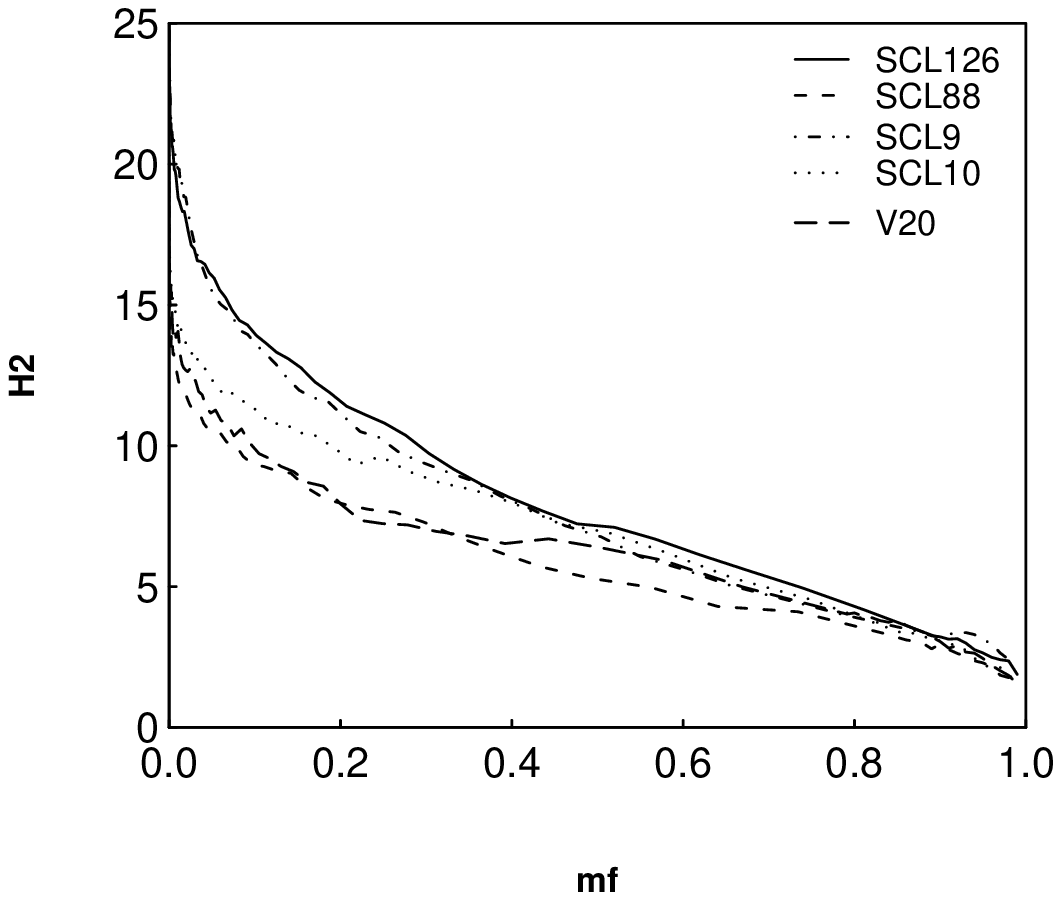}}
\resizebox{0.28\textwidth}{!}{\includegraphics*{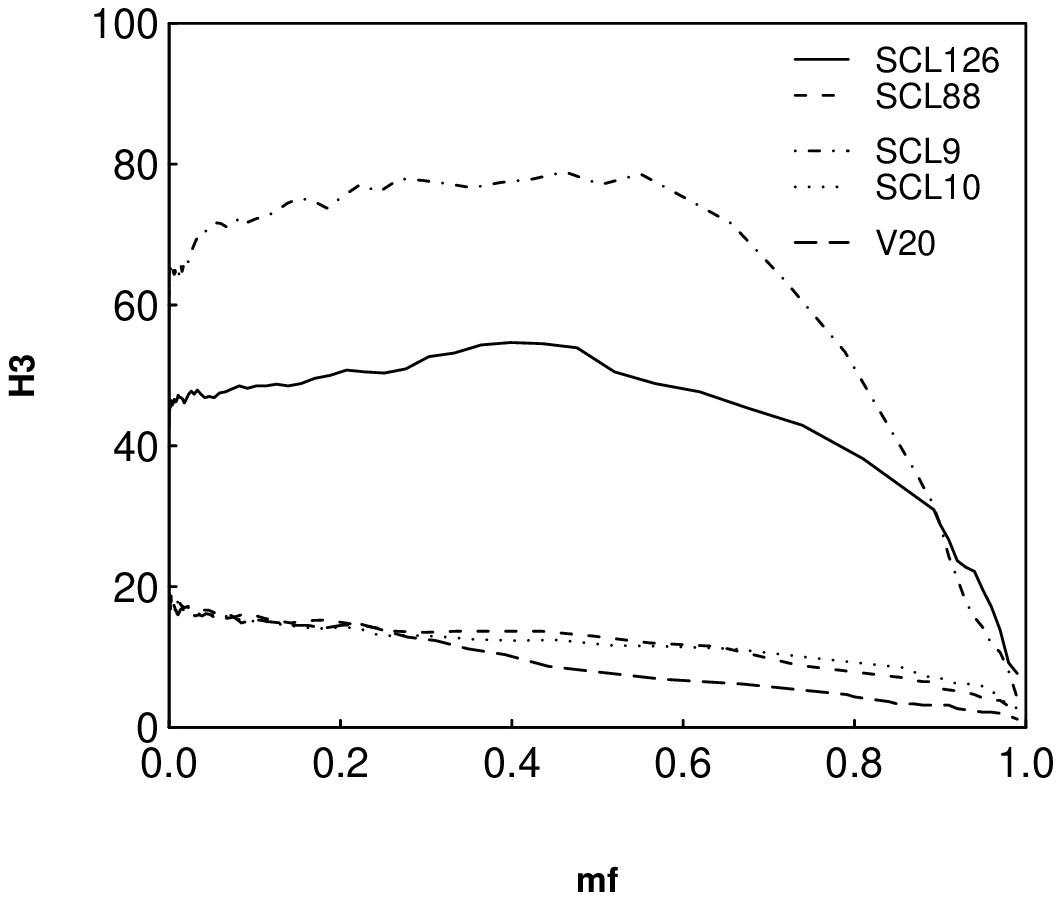}}
\hspace*{2mm}\\
\resizebox{0.28\textwidth}{!}{\includegraphics*{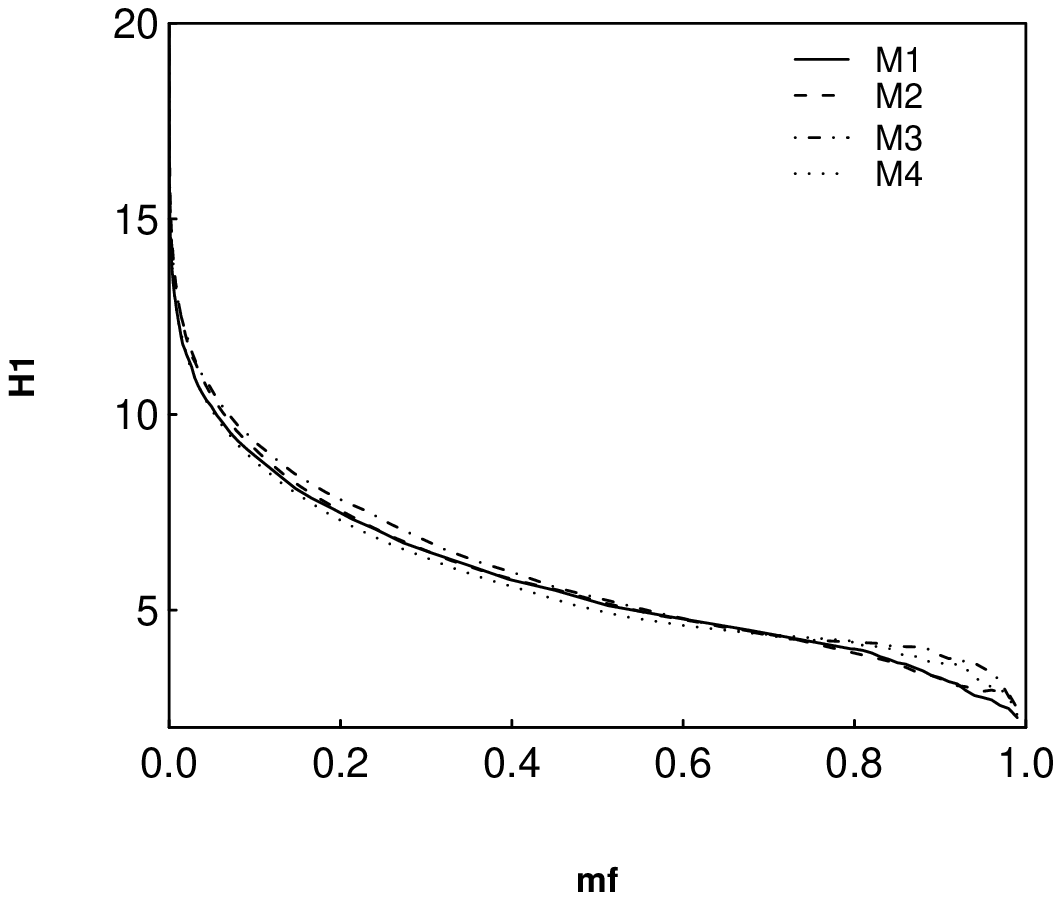}}
\resizebox{0.28\textwidth}{!}{\includegraphics*{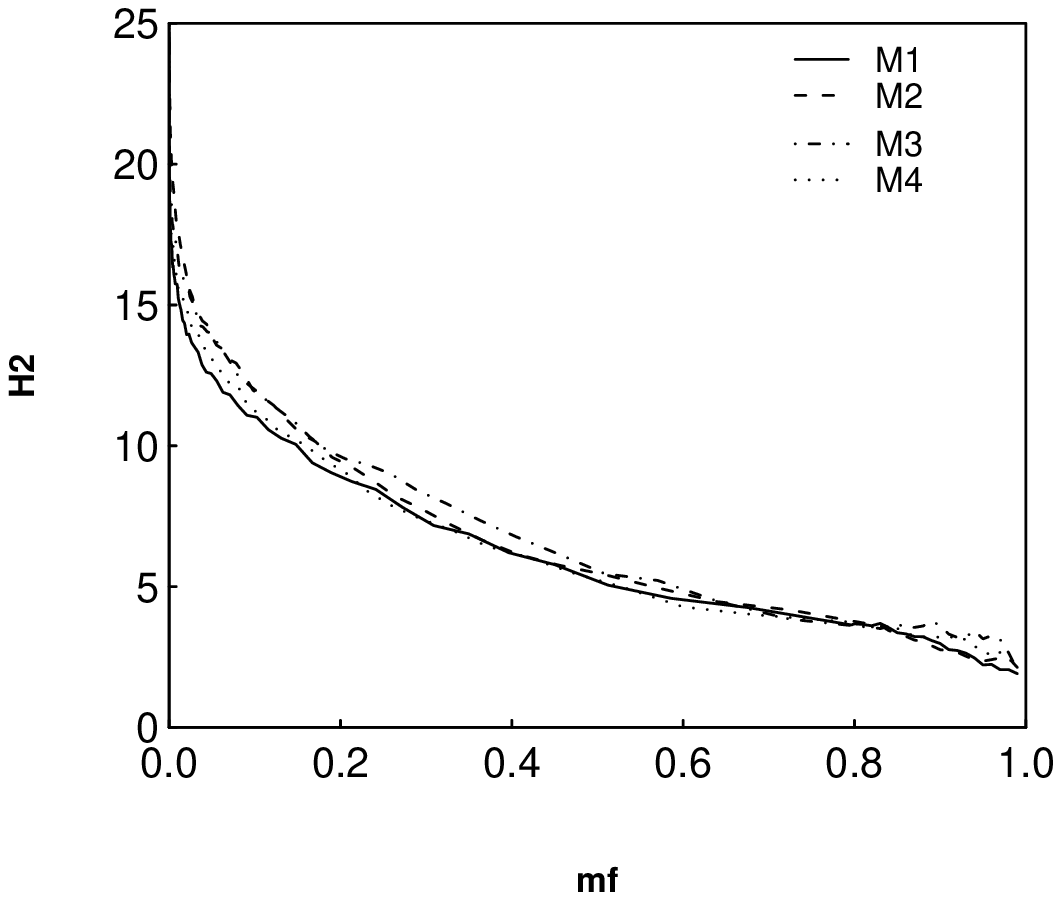}}
\resizebox{0.28\textwidth}{!}{\includegraphics*{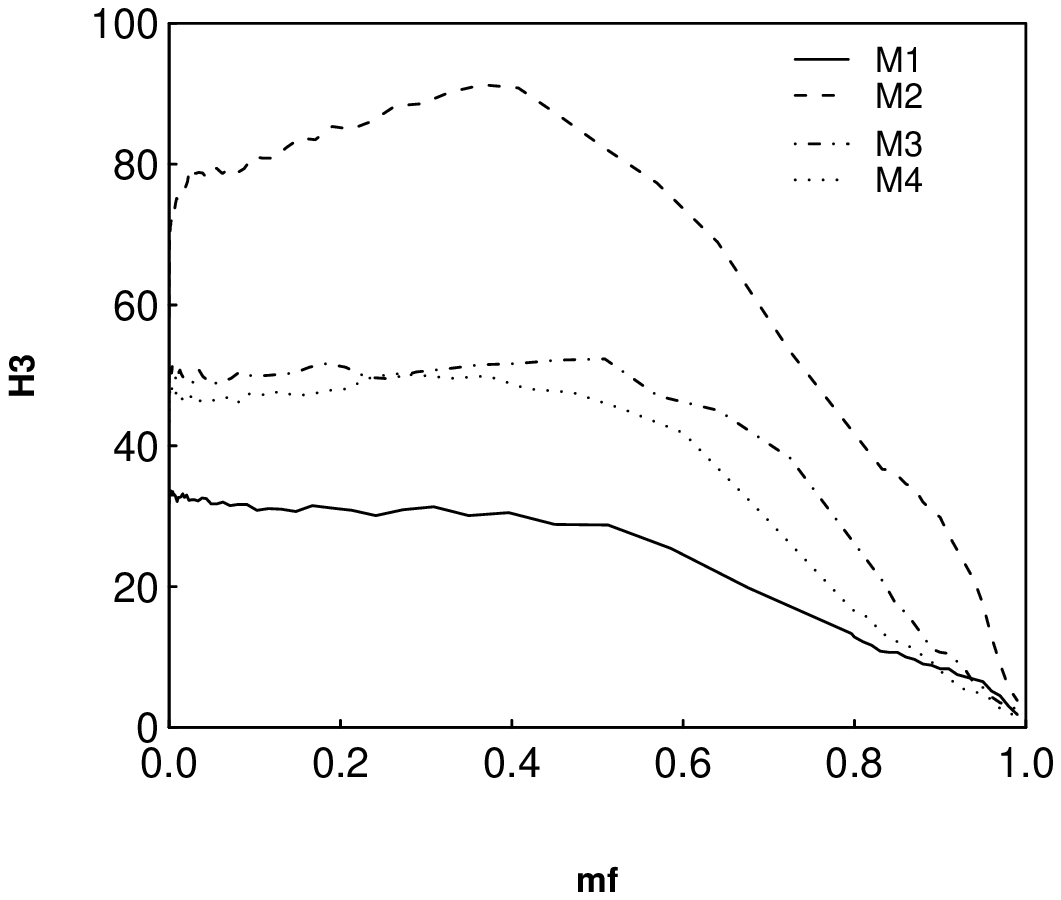}}
\caption{Shapefinders $H_{1}$ (thickness), $H_{2}$ (breadth), $H_{3}$ 
(length) (in \Mpc) for the observed superclusters  (upper panels) and for 
the simulated  superclusters (lower panels) 
versus the mass fraction $m_f$.  
}
\label{fig:h13}
\end{figure*}

\begin{figure*}[ht]
\centering
\resizebox{0.28\textwidth}{!}{\includegraphics*{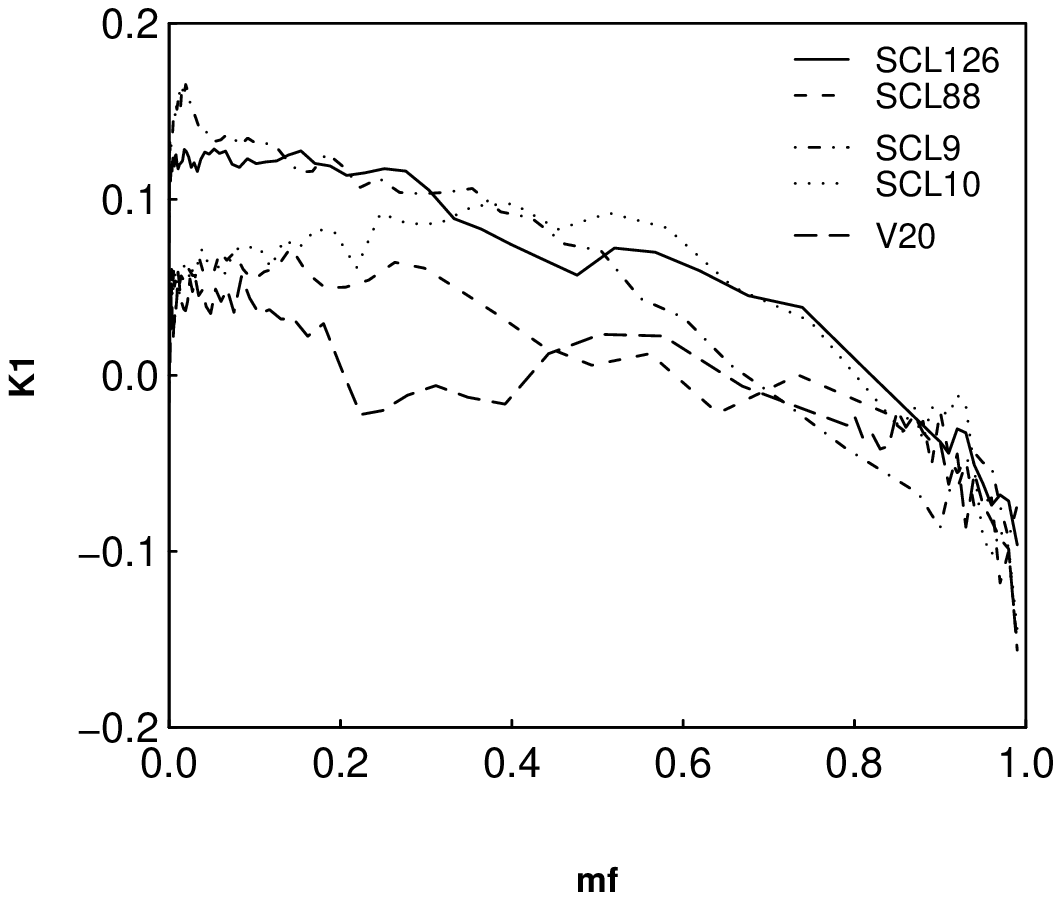}}
\resizebox{0.28\textwidth}{!}{\includegraphics*{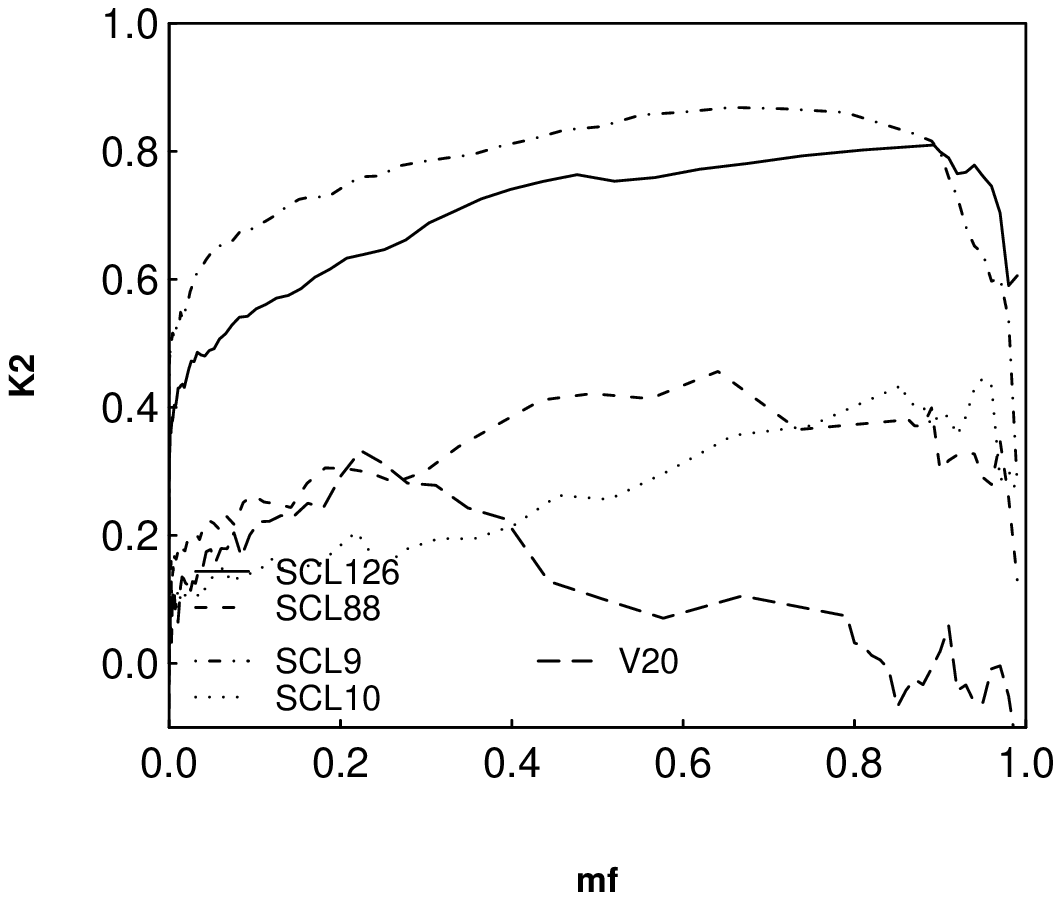}}
\resizebox{0.28\textwidth}{!}{\includegraphics*{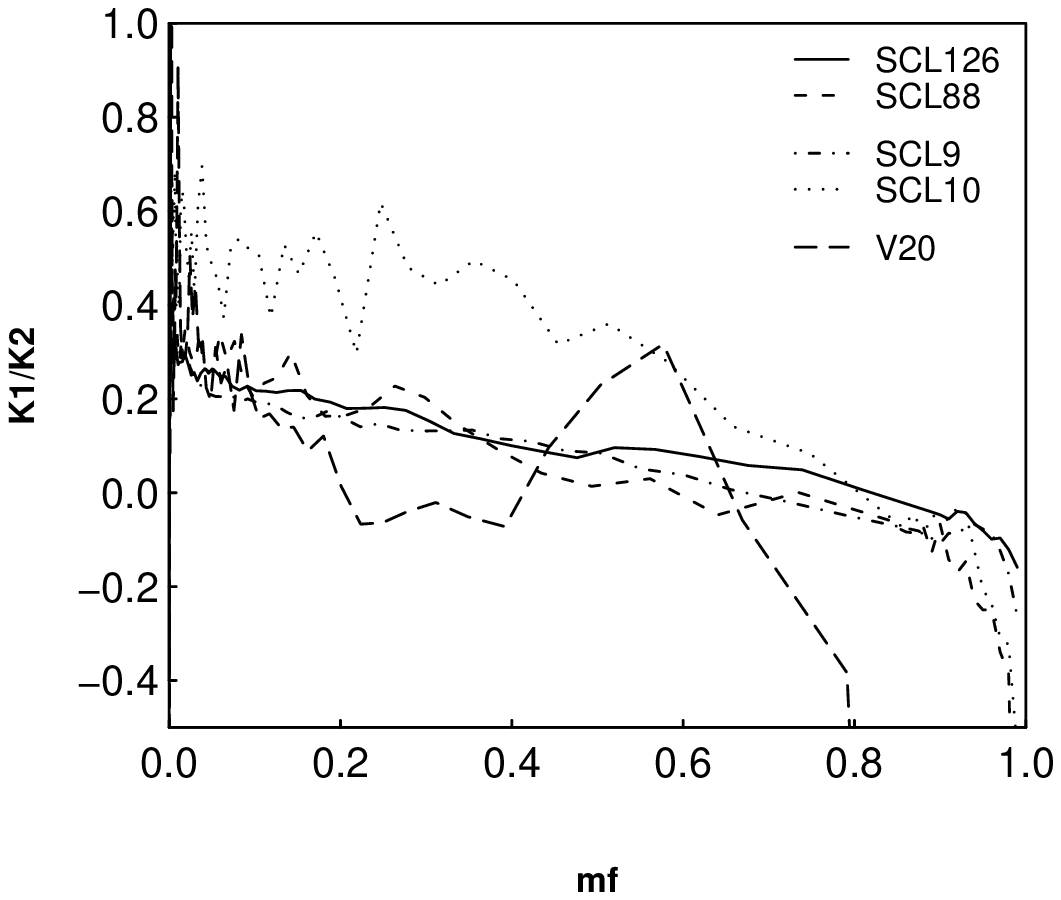}}
\hspace*{2mm}\\
\resizebox{0.28\textwidth}{!}{\includegraphics*{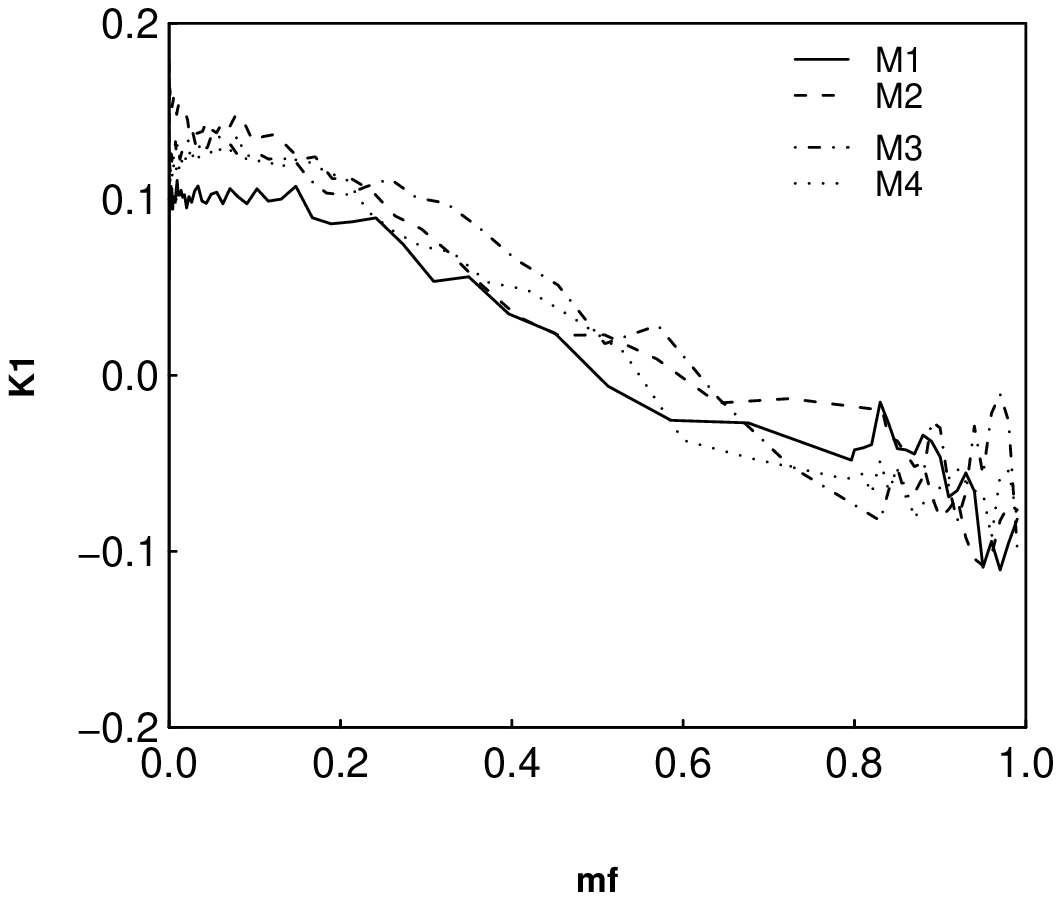}}
\resizebox{0.28\textwidth}{!}{\includegraphics*{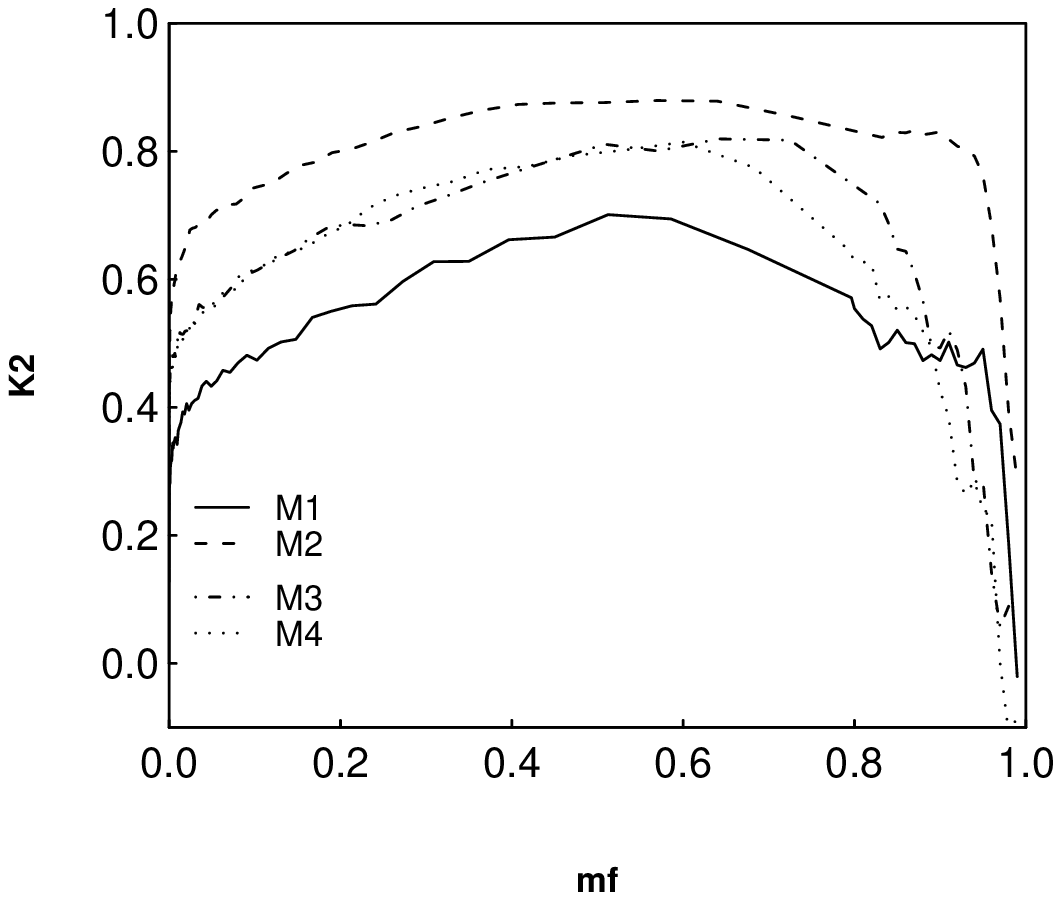}}
\resizebox{0.28\textwidth}{!}{\includegraphics*{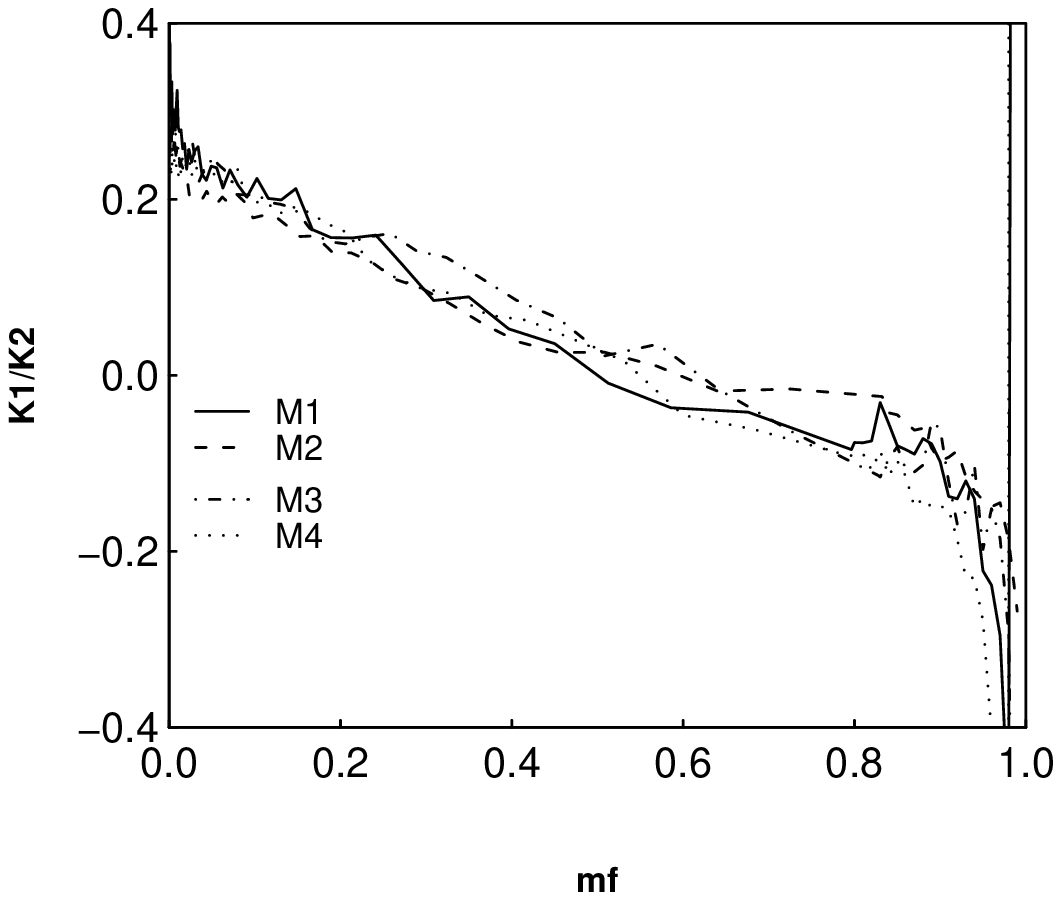}}
\caption{Shapefinders $K_1$ (planarity) and $K_2$ (filamentarity)
for the observed superclusters  (upper panels) and for 
the simulated  superclusters (lower panels) 
versus the mass fraction $m_f$. 
}
\label{fig:k12}
\end{figure*}

We present the results of our calculations of Minkowski functionals and
shapefinders in Table~\ref{tab:2} and in Figures~\ref{fig:mf4k1k2},
\ref{fig:h13} and \ref{fig:k12}. The original luminosity density field, used
to delineate superclusters (Papers I, II), was calculated using all galaxies.
For morphological study we have to use
volume-limited galaxy samples; this makes our results insensitive to
selection corrections. For this reason the density field had to be
recalculated. We used for that a kernel estimator with a $B_3$ box spline as
the smoothing kernel, with the total extent of 16~\Mpc\ (for a detailed
description see Appendix and Saar et al. \cite{saar06}). This kernel covers
exactly the 16~\Mpc\ extent of the Epanechnikov kernel, used to obtain the
original density field, but is smoother and resolves better density field
details (its effective width is about 8~\Mpc). As the argument labeling 
the isodensity surfaces, we chose the mass
fraction $m_f$ -- the ratio of the mass in regions with density {\em lower}
than the density at the surface, to the total mass of the supercluster. When
this ratio runs from 0 to 1, the iso-surfaces move from the outer limiting
boundary into the center of the supercluster, i.e. the fraction $m_f=0$
corresponds to the whole supercluster, and $m_f=1$ to its highest density
peak.

In Table~\ref{tab:2} we give for all superclusters the values of the Minkowski
functionals and shapefinders for two mass fraction values: $m_f = 0.0$, which
corresponds to the whole supercluster, and the value of $m_f$, at which the
fourth Minkowski functional $V_{3}$ has a maximum (except for SCL126, see
below) -- this shows the maximum number of isolated cores (clumps) that the
supercluster  breaks into. At lower densities (mass fractions) these clumps
are joined together, and at higher densities they start to disappear, when the
density levels get higher than their maximum density.

{\scriptsize
\begin{table*}[ht]
\caption{The Minkowski Functionals and shapefinders for rich superclusters. }
\begin{tabular}{rrrrrrrrrrrrr} 
\hline 
 ID     &$m_f$&   $V_{0}$  &  $V_{1}$ & $V_{2}$ & $V_{3}$ & $H_{1}$ (T)& $H_{2}$ (B)
 & $H_{3}$ (L)&$K_{1}$(P) & $K_{2}$(F) & $K_{1}$/$K_{2}$ \\
\hline 
SCL126  &0.0&  1.27e5  & 2566.0  & 56.9   & 1  & 24.67  & 28.71  & 42.67  &  0.08   & 0.20 &  0.39\\ 
        &0.89&   1462  &  210.2  & 41.2   & 9  &  3.48  &  3.25  & 30.91  & -0.03   & 0.81 & -0.04\\ 
\\          
SCL88   &0.0 & 2.84e4  & 743.10  & 24.44  & 1  & 19.11  & 19.36  & 18.33  &  0.01   & -0.03 & -2.36e-01\\
        &0.65 &   927  & 103.40  & 15.33  & 3  &  4.48  &  4.29  & 11.50  & -0.02   &  0.46 & -4.71e-02\\
\\          
SCL9    &0.0 & 1.75e5  & 3591.0 &  78.11  &  1 & 24.42  & 29.27  & 58.58  &  0.09   & 0.33  &  0.27\\ 
        &0.79&   3855  &  445.3 &  71.11  & 15 &  4.33  &  3.99  & 53.33  & -0.04   & 0.86  & -0.05\\ 
\\          
SCL10   &0.0 & 3.72e4    & 885.80 & 27.11  & 1 & 20.97  & 20.80  & 20.33  & -0.004  & -0.01 &   0.35\\ 
        &0.86&   425     &  59.56 & 11.33  & 3 &  3.57  &  3.35  &  8.50  & -0.03   &  0.43 &  -0.07\\ 
\\
V20   &0.0& 3.05e4    & 736.90  & 23.33  & 1  & 20.72    & 20.11 & 17.50   & -0.02  & -0.07 &  0.22\\ 
        &0.43&  1589    & 121.60  & 11.56  & 2  &  6.53    &  6.70 &  8.67   &  0.01  &  0.13 &  0.10\\
\\             
M1      &0.0& 7.29e4  & 1649.00  & 42.56   & 1 & 22.09  & 24.67  & 31.92  &  0.06  &  0.13  &  0.43\\ 
        &0.59&  2354  &  244.00  & 33.89   & 5 &  4.82  &  4.58  & 25.42  & -0.03  &  0.69  & -0.04\\ 
\\          
M2      &0.0& 1.82e5    & 3772.0  &  77.78 &  1    & 24.18  & 30.87 & 58.34 &  0.12  & 0.31 &  0.40\\ 
        &0.65&   5930   &  644.0  &  91.89 & 15    &  4.60  &  4.46 & 68.92 & -0.02  & 0.88 & -0.018\\
\\          
M3      &0.0& 1.19e5    & 2624.0  & 63.67  & 1  & 22.64    & 26.24  & 47.75 &  0.07  & 0.29  &  0.25\\ 
        &0.65&   3860   &  420.9  & 60.00  & 9  &  4.59    &  4.47  & 45.00 & -0.01  & 0.82  & -0.02\\ 
\\          
M4      &0.0& 1.06e5    & 2381.00 & 58.00  & 1  & 22.26   & 26.13   & 43.50 &  0.08  &  0.25  &    0.32\\
        &0.60&   3454   &  374.90 & 55.78  & 9  &  4.61   &  4.28   & 41.84  & -0.04 &  0.81  &   -0.05\\ 

\label{tab:2}     
\end{tabular}  

The columns in the Table are as follows:
\noindent column 1: Supercluster ID,
\noindent column 2: mass fraction, $m_f$,
\noindent columns 3--6: Minkowski functionals $V_{0}$ -- $V_{3}$, 
\noindent $V_0$ in (\Mpc)$^3$, $V_1$ in (\Mpc)$^2$,  $V_2$ in \Mpc,  
\noindent columns 7--9: shapefinders $H_1$ (thickness), $H_2$ (breadth)
and $H_3$ (length), in \Mpc, 
\noindent columns 10--12: shapefinders $K_1$ (planarity), $K_2$
(filamentarity) and their ratio, $K_1/K_2$. 

\end{table*}            
}

At small mass fractions the iso-density surface includes the whole
supercluster. Thus volumes and areas of superclusters ($V_{0}$ and
$V_{1}$, Table~\ref{tab:2}) are large. As we move to higher mass
fractions, the iso-density surfaces include only higher density parts
of superclusters, and their volumes and areas get smaller. At very
high mass fractions only the highest density clumps in superclusters
give their contribution to the supercluster. Individual high density
regions in a supercluster, which at low mass fraction are joined
together into one system, began to separate from each other, and the
value of the fourth Minkowski functional ($V_{3}$) increases. At a
certain density contrast (mass fraction) $V_{3}$ has a maximum showing
the largest number of isolated clumps in a given supercluster at the
spatial resolution determined by the smoothing kernel. At still higher
density contrasts only the highest density peaks contribute to the
supercluster.

Figure~\ref{fig:mf4k1k2} (left panels) shows the fourth Minkowski
functional $V_{3}$ for the most massive real and simulated
superclusters. 
The $V_{3}$ curve for the supercluster SCL9 in the 
left upper panel shows a characteristic behaviour. 
At the mass fraction value of about 0.2, the
value of $V_{3}$ for SCL9 begins to increase and reaches a maximum
value at the mass fraction $m_f \approx 0.7 $. Then the values of
$V_{3}$ begin to decrease. This indicates that the overall morphology
of the supercluster SCL9 is clumpy; this supercluster consists of a
large number of clumps or cores connected by relatively thin
filaments, in which the density of galaxies is too low to contribute
to the supercluster, starting at certain mass fraction values. The
maximum value of the fourth Minkowski functional $V_{3}$ shows that
the supercluster SCL9 has the largest number of isolated clumps in it,
although this supercluster is only partly covered by our sample -- the
value of $V_3$ for the whole supercluster may be twice as large as our
present calculations show. This supercluster is the largest and
richest of observed superclusters in our present sample, with the
largest size and volume.

The second richest and largest supercluster among the observed
superclusters is the supercluster SCL126.  Figure~\ref{fig:mf4k1k2}
shows that the $V_{3}$ curve for SCL126 has a shape which is 
quite different from that for SCL9. At a mass
fraction $m_f \approx 0.4 $ the values of $V_{3}$ increase rapidly,
has a peak, decreases again and has another peak at a mass fraction of
about $m_f \approx 0.85$. This behaviour of the $V_{3}$ curve
indicates that the overall morphology of the supercluster SCL126 is
rather homogeneous, which is characteristic to a rich filament with
several branches (see also Sect. 3.3). Interestingly, the $V_{3}$
curve for the supercluster SCL126 shows several peaks at a high mass
fraction, $m_f > 0.95$. This indicates the presence of a very high
density core region with several individual clumps in it -- this is
the main core region of the supercluster with several Abell clusters,
which are also X-ray clusters (Einasto et al. \cite{e03d}). In other
superclusters we do not see such a high density and very compact
core. An example of a supercluster with a high density
core is the simulated supercluster M2, but in this supercluster the
peaks at high mass fraction in the $V_{3}$ curve appear at mass fractions 
$m_f < 0.95$.

In Figure~\ref{fig:mf4k1k2} the observed superclusters separate clearly 
into two different classes. The superclusters SCL88 and SCL10 have much 
smaller numbers of  density peaks than the superclusters SCL126 and 
SCL9. This may be partly explained by the incompleteness of the SCL10
and SCL88, which are 
cut by the survey boundaries. In the case of 
the Virgo supercluster V20 the maximum value of the fourth Minkowski 
functional is only 2, describing a compact supercluster.

The shapes of the $V_{3}$ curves for simulated superclusters in
Figure~\ref{fig:mf4k1k2} and their maximum values show large
variations.  In the case of the supercluster M2 the $V_{3}$ curve
shows a rapid increase at the mass fraction $m_f \approx 0.4$, and
three maxima, one of them at 
$m_f \approx 0.9$. This shows that this
supercluster is very clumpy. The maximum value of $V_{3}$ is
comparable to that for the observed supercluster SCL9, and the
presence of a peak at high values of $m_f$ is comparable to that for
SCL126. However, the mass fraction at which the peak occurs is lower
than in the case of SCL126.

The shapes of the $V_{3}$ curves for simulated  superclusters M3 and M4 
resemble those for the observed supercluster SCL9; however, the maximum 
values of $V_{3}$ are less than 10, showing that the number of isolated 
cores or clumps in these simulated superclusters is smaller than in 
SCL9. The simulated supercluster M1 has the smallest maximum value of 
$V_{3}$ among simulated superclusters showing that this supercluster is 
less clumpy than other simulated superclusters. The number is isolated 
clumps in this supercluster is still larger than that for the observed 
superclusters SCL10 and SCL88.

We have determined the number of density field clusters in superclusters
(Paper I). Tables~\ref{tab:1} and ~\ref{tab:2} show that in all
superclusters except 
SCL88 and M4 the maximum value of $V_{3}$ is about half the
number of density field clusters indicating that typically high
density cores in superclusters contain two density field clusters. 
In superclusters SCL88 and M4 these values are equal.

The value of $V_{3}$ for SCL9 and M2 is negative at the mass fraction $m_f
\approx 0$, indicating that there are holes through the supercluster.

Next we analyze the shapefinders $H_{1}$-$H_{3}$ for
superclusters. These quantities have dimensions of length, and
$H_1\leq H_2\leq H_3$ in the case of a convex body (e.g., triaxial
ellipsoid). Therefore, they can be used to study the dimensions of
superclusters.  The shapefinder $H_{1}$ is the smallest and
characterizes the thickness of superclusters. The shapefinder $H_{2}$
as an intermediate one is an analogy of the breadth of a
supercluster. The breadth is calculated as $H_2 = S/C$, it contains
information about both the area and curvature of an isodensity
surface. The shapefinder $H_{3}$ is the longest and describes the
length of the superclusters. Of course, this is not the real length of
the supercluster, but a measure of the integrated curvature of the
surface which may become very large for irregularly shaped and curved
surfaces.

In Figure~\ref{fig:h13} (upper panels) we plot the shapefinders
$H_{1}$ to $H_{3}$ for the richest superclusters. This figure shows
that the extension of the superclusters as measured by the
shapefinders $H_{1}$ and $H_{2}$ is about 15 -- 20~\Mpc\ for the
complete supercluster ($m_f=0$). At higher mass fractions, $m_f
\approx 0.5$, the iso-surfaces include only higher density parts of
superclusters, $H_{1}$ and $H_{2}$ are less than 10~\Mpc, i.e. the
supercluster centers are still typical 3-dimensional objects. At those
mass fractions which correspond to the maximum value of $V_{3}$ (the
core regions of superclusters) $H_{1}$ and $H_{2}$ are of about
5~\Mpc. The scatter of the shapefinder $H_{2}$ for observed
superclusters is larger than that of $H_{1}$ showing the influence of
a different number of substructures (isolated clumps or cores) in
these systems.

The shapefinder $H_{3}$ differs strongly for the four observed
superclusters.  For the whole superclusters ($m_f=0$) the values of
$H_{3}$ are about 20\Mpc\ for poorer superclusters and 40 -- 60~\Mpc\
for the two richest superclusters, SCL126 and SCL9. We note that for
the richest two superclusters at mean mass fractions $m_f \approx 0.5$
the value of $H_{3}$ is larger than at $m_f=0$, reaching a maximum
value of about 60 -- 80~\Mpc. This shows their complicated structure
with subsystems of large curvature. The value of $H_{3}$ is the
largest in the case of the supercluster SCL9, which is the longest
supercluster with the largest number of isolated clumps or cores in
it. In the case of other observed superclusters the value of $H_{3}$
decreases when we increase the mass fraction and move into the central
parts of the superclusters.  This is additional evidence that these
superclusters are less clumpy than SCL9 and SCL126.  At mass fractions
$m_f \approx 0.5$ the value of $H_{3}$ for these superclusters is less
than 20~\Mpc.

Figure~\ref{fig:h13} (lower panels) shows the shapefinders $H_{1}$-$H_{3}$ 
for simulated superclusters. We see that the shapefinder $H_{1}$ 
(thickness)  for the simulated superclusters has values close to those 
for observed superclusters, but with a much smaller scatter. The breadths 
of the simulated superclusters ($H_{2}$) have values intermediate between 
those for the observed superclusters SCL9 and SCL126, and for other observed 
superclusters. Again, the scatter of these values is very small. 
Therefore, the shapefinders $H_{1}$ and $H_{2}$ for the model superclusters 
show an astonishing universality.

The shapefinder $H_{3}$ (length) shows a rather different
picture. 
The curve for the simulated supercluster M2 is rather
similar to that for the observed supercluster SCL9; $H_{3}$ for the
superclusters M3 and M4 are close to that for SCL126. The length of
the shortest simulated supercluster M1 is still larger than the length
of the shortest observed superclusters, SCL10, SCL88, and the Local
supercluster. The large values of $H_{3}$ at intermediate mass
fractions, $m_f \approx 0.5$ (30 and 90~\Mpc) indicate the presence of
substructures in superclusters with high values of the curvature $C$.

Next we study the shapefinders $K_{1}$ and $K_{2}$ for the richest
superclusters (Figure~\ref{fig:k12}). $K_{1}$ is defined by the
thickness $H_{1}$ and the breadth $H_{2}$, this characterizes the
planarity of the superclusters; $K_{2}$ is calculated from the breadth
$H_{2}$ and the length $H_{3}$ and this parameter characterizes the
filamentarity of superclusters.

In the upper panels of Fig.~\ref{fig:k12} we present the planarity and
the filamentarity for observed superclusters. The values of the
planarity $K_1$ for the full superclusters (the mass fraction $m_f=0$)
are 0.10--0.15 for the richest superclusters, SCL126 and SCL9, and
about 0.05 for other observed superclusters.  As the mass fraction
increases and only the higher density parts contribute to
superclusters, the values of the planarity $K_1$ start to decrease.
For the supercluster SCL126 the $K_1$ curve has a small minimum at the
mass fraction $m_f \approx 0.4 $; this is the mass fraction value at
which the value of the fourth Minkowski functional $V_{3}$ starts to
increase. In superclusters the value of $K_1$ for the core regions
becomes negative. This shows that at very high mass fractions, which
include only the central regions of superclusters, the isodensity
surfaces have complex shapes, different from the heuristic
classification based on convex ellipsoids 
as described above.

The values of the filamentarity for observed superclusters, $K_2$,
have much larger scatter than the values of the planarity, $K_1$.  For
the richest superclusters, SCL126 and SCL9, $K_2 \approx 0.2 - 0.3$,
the other observed superclusters have smaller $K_2$ (at the mass fraction
$m_f = 0$).  The central parts 
of these superclusters (at high values of mass fractions) 
are more filamentarity than the full superclusters. The
Virgo supercluster V20 has a different shape, for this supercluster the
the filamentarity $K2$ decreases for $mf < 0.2$.

In the case of simulated superclusters, the filamentarities $K_2$ have
a smaller scatter than those for the observed superclusters. The
simulated supercluster M2 has the largest value of $K_2$ in our
sample, $K_2 \approx 0.6$, other simulated superclusters have $K_2
\approx 0.3-0.5$, similar to the observed superclusters SCL126 and
SCL9 (for the whole supercluster).

In earlier studies the shapes of superclusters have been characterized
using the ratio $K_1$/$K_2$ for the full superclusters (Basilakos et
al.  \cite{bpr01}; Basilakos \cite{bas03}). We plot this ratio for the
whole mass fraction interval in Figure~\ref{fig:k12}, right
panels. This Figure shows that the ratios $K_1$/$K_2$ for the observed
and simulated superclusters are rather similar, having values of about
0.25--0.4. This shows a high degree of filamentarity in the case of
the whole superclusters. The ratios $K_1$/$K_2$ slowly decrease, as
we increase the mass fraction and move to central regions of
superclusters.  Exceptions are the observed supercluster SCL10 with
the highest values of 
$K_1$/$K_2$ at intermediate mass fractions, and the Virgo supercluster V20 
for which this ratio changes strongly. 

The information about the shapes of superclusters can be best
described by their morphological signature, the path in the
shapefinder $K_1$-$K_2$ plane for varying $m_f$
(Figure~\ref{fig:mf4k1k2}, right panels), both for the observed and
model superclusters. To show which part of the shape plane corresponds
to the whole supercluster, we mark with triangles the values of
$K_1,K_2$ at the mass fraction $m_f = 0$ for the superclusters SCL126
and M1. We also mark with circles the values of $K_1,K_2$ at the mass
fraction corresponding to the maximum value of the fourth Minkowski
functional $V_{3}$ (Table~\ref{tab:2}).  As explained in Appendix B,
we restrict the $m_f$ from below, starting the curves from
$m_f=0.01$. This is done to eliminate the influence of the slight
non-isotropy of the $B_3$ kernel at low densities.

In the shapefinder plane $K_1$-$K_2$, the observed superclusters
SCL126 and SCL9, and the simulated superclusters have similar
trajectories. As we change the mass fraction, the $K_1$--$K_2$
shapefinder path moves 
from low $K_1$ and $K_2$ values (this corresponds
to the whole supercluster and low mass fractions) to the upper left
region with higher $K_2$ and smaller $K_1$ 
(high mass fractions, the core of the supercluster). At first,
as the mass fraction increases, the value of the shapefinder $K_1$
(the indicator of planarity) almost does not change, but the value of
the shapefinder $K_2$ (the indicator of filamentarity) increases, in
accordance to what we saw in Figure~\ref{fig:k12}. At a certain mass
fraction the value of the shapefinder $K_2$ reaches it's maximum
value. As we still increase the value of the mass fraction (and move
to higher densities, into the cores of superclusters), the value of
the shapefinder $K_2$ changes a little, but the value of the
shapefinder $K_1$ decreases.

We see that the richest superclusters have a distinct signature in the
shapefinder $K_1$-$K_2$ plane -- a characteristic curve which
describes the typical morphology of superclusters. This signature is
characterized by an rising path with a small fixed positive $K_1$, a
plateau at the maximum value of $K_2$, and a descending branch at a
small fixed negative value of $K_1$.  In Appendix we shall show that
this curve is characteristic to multi-branching filaments.

The large scatter of the different curves in the $K_1$--$K_2$ plane is
remarkable. 
Among the observed superclusters, we find two types of behaviour. 
The scatter of the trajectories for the model
superclusters is smaller.

So, in summary, superclusters are extended 3-dimensional objects
composed of multi-branched filaments. The clumpiness of superclusters
can be quantified by the fourth Minkowski functional $V_3$ which
determines the number of isolated cores or clumps in
superclusters. The shape of the $V_{3}$ curve gives us information
about the overall morphology of the superclusters (a rich filament
with several branches in the case of SCL126, clumpy in the case of
other superclusters). In the $K_1$--$K_2$ shapefinder plane the
morphology of superclusters is described by a curve (morphological
signature), which is characteristic to multi-branching filaments. In
case of the Virgo supercluster we see how the Minkowski functionals
and shapefinders describe a compact supercluster with one central
cluster and accompanying filaments, where poorer clusters and groups
of galaxies reside.  The morphological signature for the Virgo
supercluster is characteristic to a spider -- a supercluster with one
central body, surrounded by filaments (see Appendix).

Using the data in Table~\ref{tab:2} we can estimate that in high
density regions of superclusters (at mass fractions where the value of
the fourth Minkowski functional has a maximum) the density of galaxies
is of about ten times higher than the mean density of galaxies in the
same supercluster.

\begin{figure*}[ht]
\centering
\resizebox{0.22\textwidth}{!}{\includegraphics*{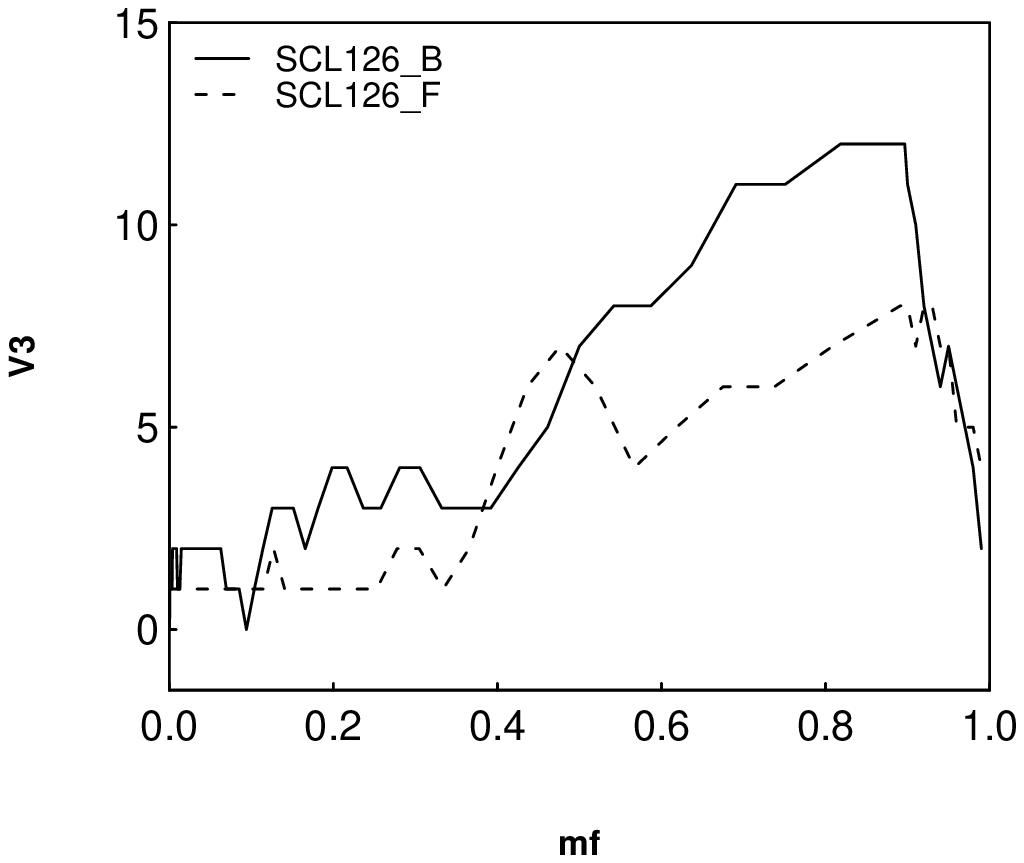}}
\resizebox{0.22\textwidth}{!}{\includegraphics*{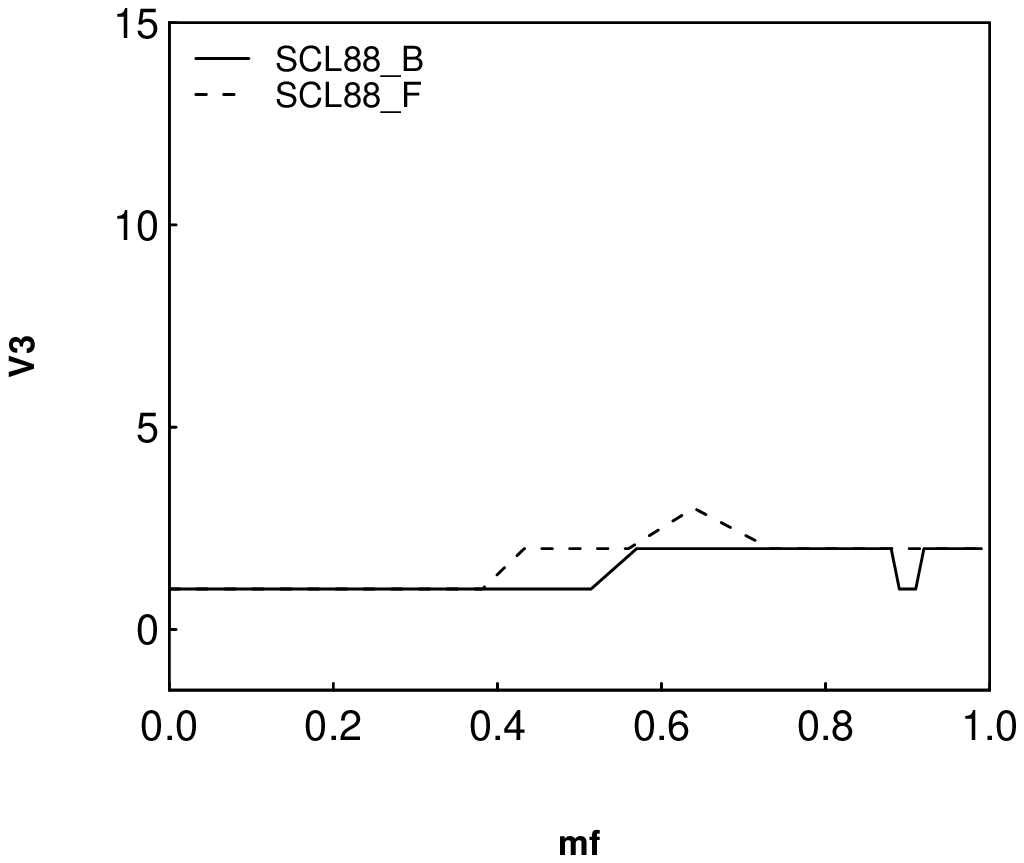}}
\resizebox{0.22\textwidth}{!}{\includegraphics*{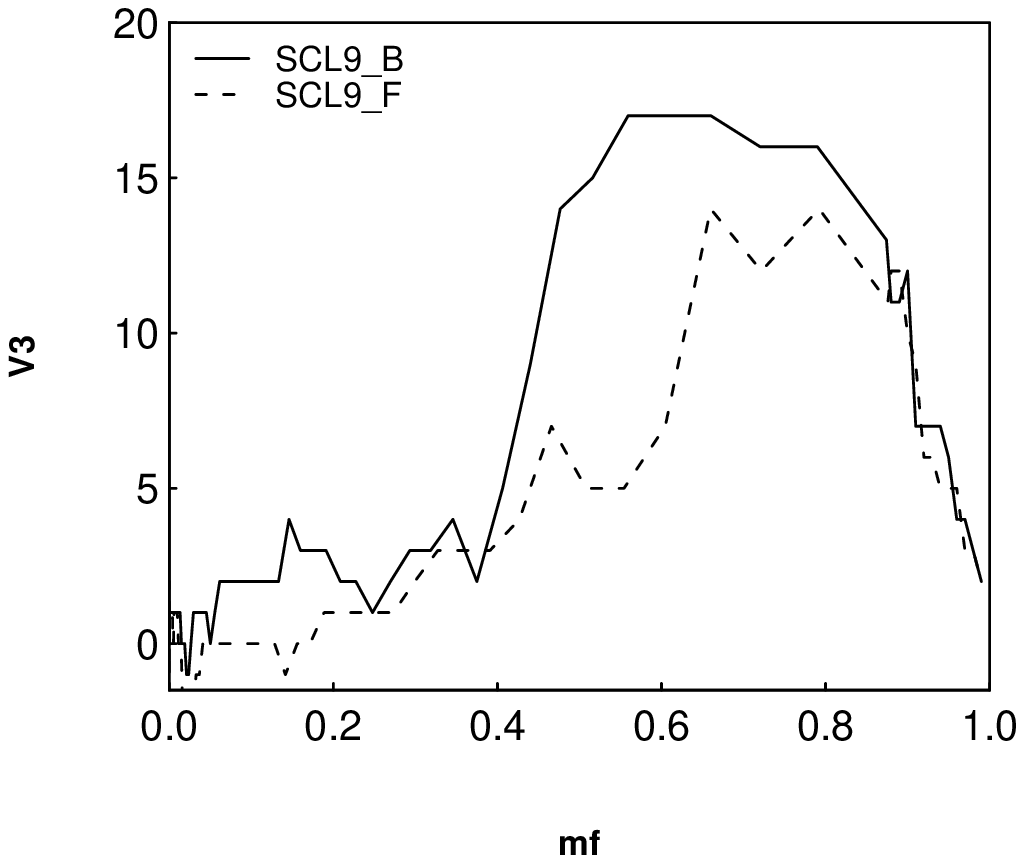}}
\resizebox{0.22\textwidth}{!}{\includegraphics*{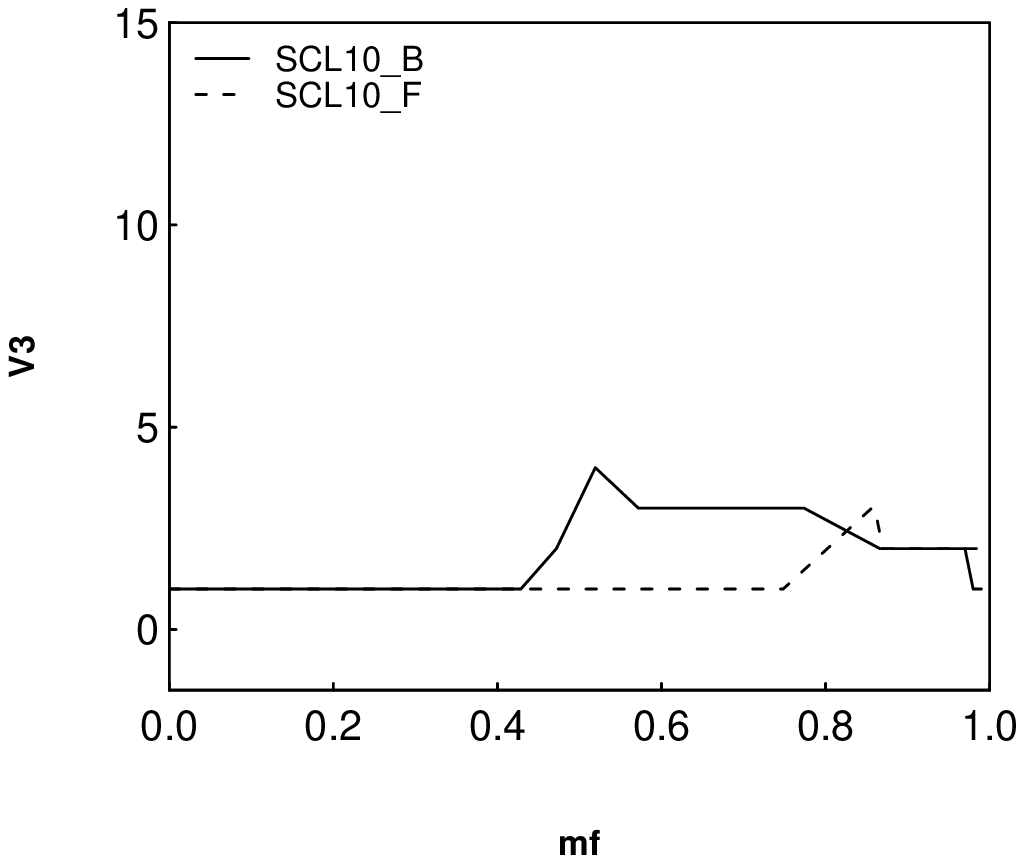}}
\hspace*{2mm}\\
\resizebox{0.22\textwidth}{!}{\includegraphics*{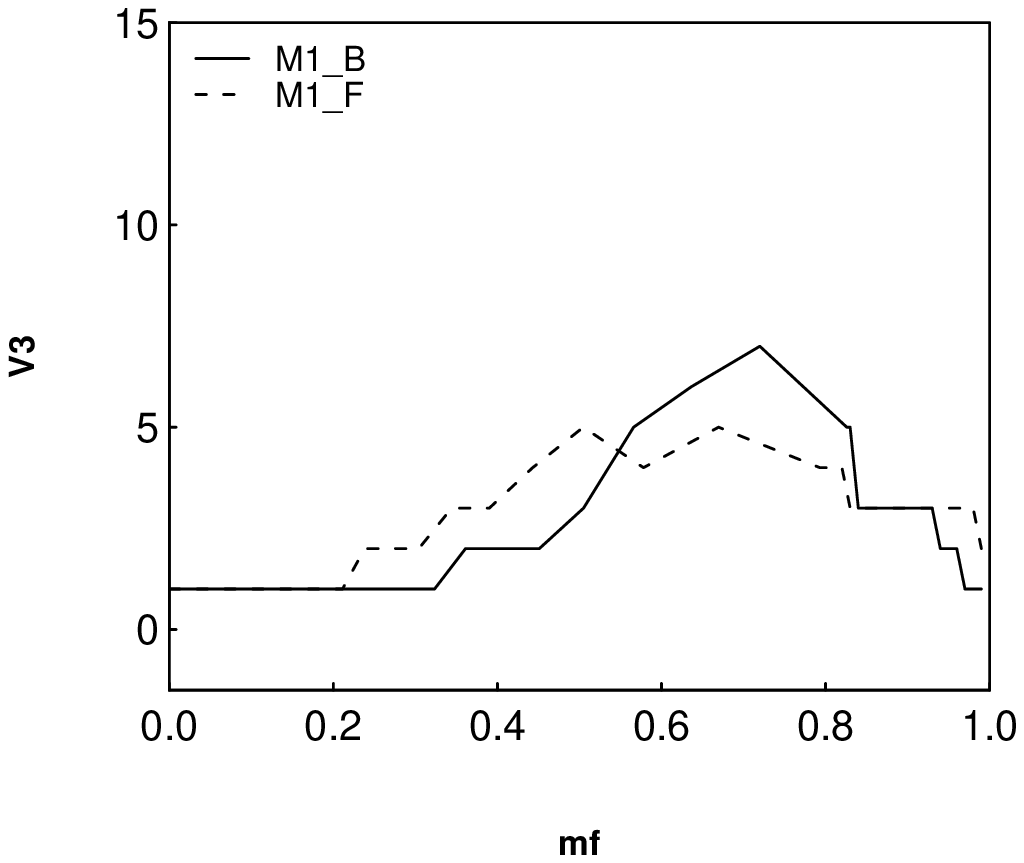}}
\resizebox{0.22\textwidth}{!}{\includegraphics*{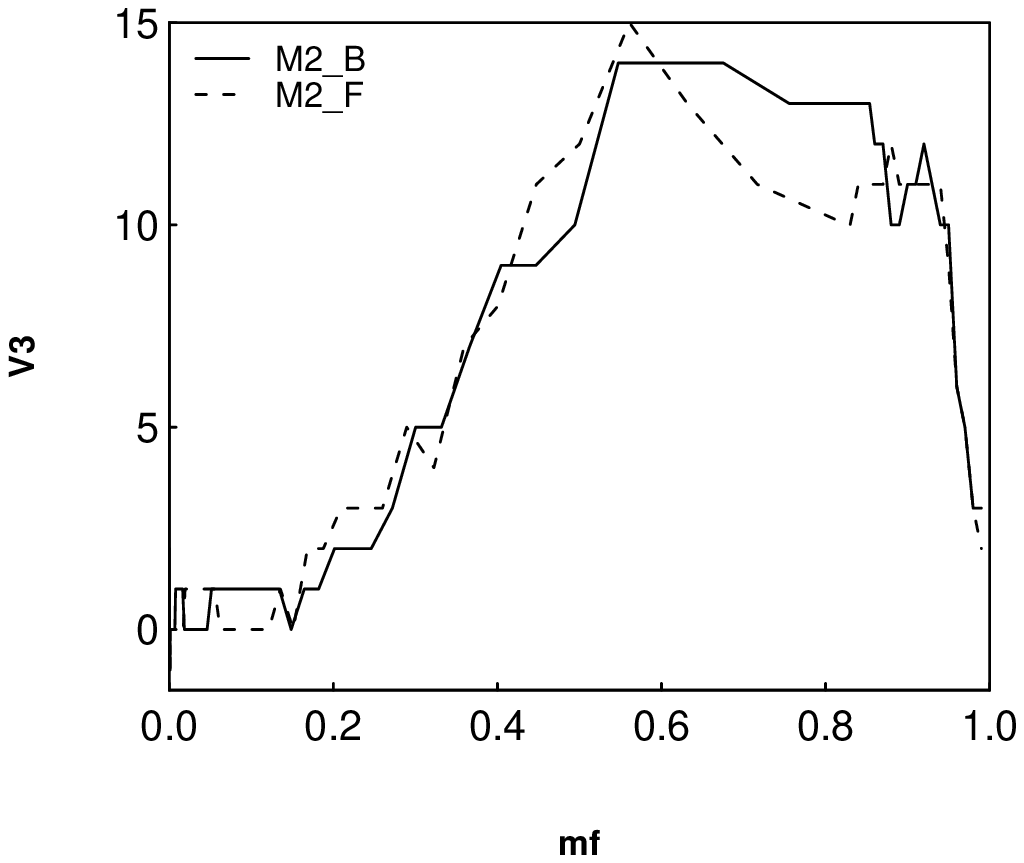}}
\resizebox{0.22\textwidth}{!}{\includegraphics*{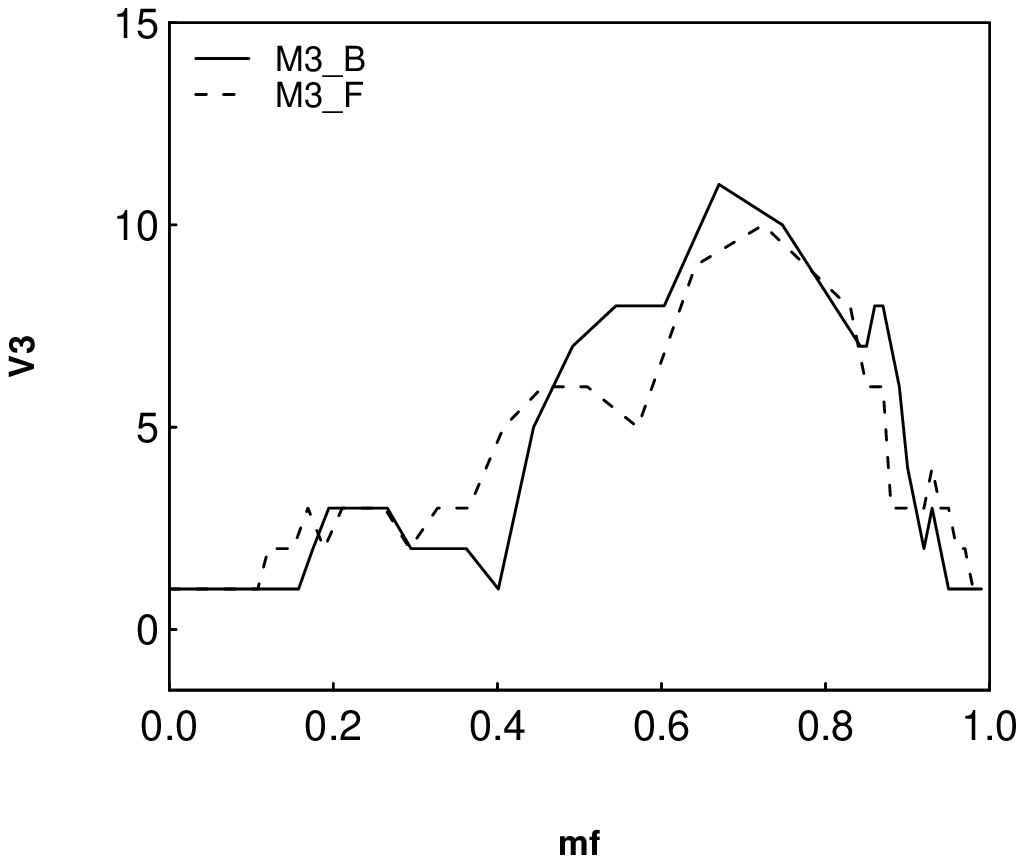}}
\resizebox{0.22\textwidth}{!}{\includegraphics*{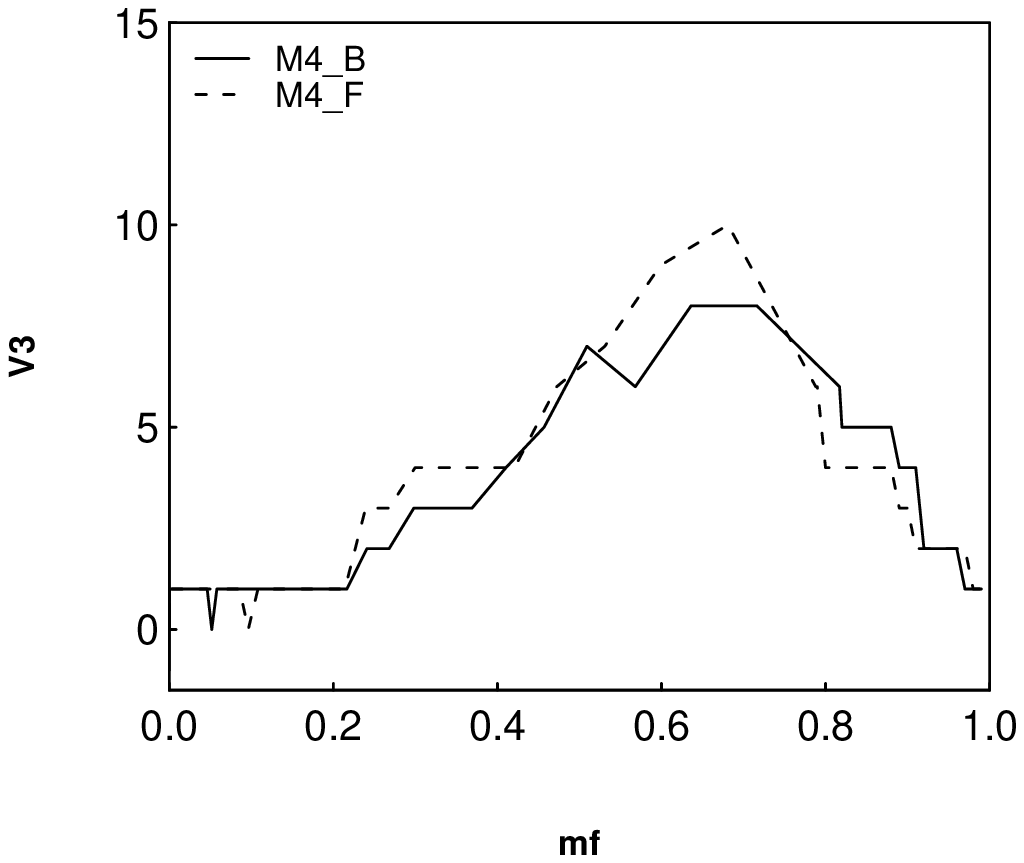}}
\caption{The fourth Minkowski functional $V_{3}$
for the bright (B, $M \leq -20.0$) and faint (F, $M > -20.0$) galaxies
of the observed superclusters (upper 
panels, from left to right: SCL126, SCL88, SCL9 and SCL10)
and for the Millennium simulation (lower panels,
from left to right: M1, M2, M3, M4).
}
\label{fig:smillbfmf3}
\end{figure*}

\subsection{Distribution of bright and faint galaxies}

As a further application of Minkowski functionals we investigate the
distribution of bright and faint galaxies in rich superclusters.  We
divide galaxies into populations of bright and faint galaxies using a
bright/faint galaxy limit $M_{bj} = -20.0$.  The same division was
used in Paper III (Einasto et al. 2007c) 
to study the distribution of bright and faint
galaxies in superclusters. Next we calculate the Minkowski functionals
separately for these two populations of galaxies, for both real and
simulated superclusters.

The fourth Minkowski functional, $V_{3}$, for the bright and faint
galaxies in the observed and simulated superclusters is shown in
Fig.~\ref{fig:smillbfmf3}. We see that there are large differences of
the fourth Minkowski functional for these populations in observed
superclusters.  This Minkowski functional characterises the clumpiness
of superclusters. In the supercluster SCL126 a high level of
clumpiness is observed over a large range of mass fractions, both for
the bright and faint galaxies.  For the bright galaxies $V_{3}$
reaches a value of about 10, while for the faint galaxies the value of
$V_{3}$ remains about 5.  In the supercluster SCL9 the clumpiness of
both the bright and faint galaxies is peaked at a rather high value of
the mass fraction, but for the bright galaxies in a broader mass
fraction range. In the superclusters SCL88 and SCL10 the clumpiness is
very low, again for galaxies of both brightness classes. The fact that
the values of $V_{3}$ for the bright galaxies are larger than for the
faint galaxies shows that the bright galaxies are located in numerous
clumps or cores while the fainter galaxies form a less clumpy
population around them.

In contrast, the values of $V_{3}$ for the bright and faint galaxies in 
simulated superclusters differ less than in the case of the observed 
superclusters. Therefore, the clumpiness of the bright and faint galaxies in 
model superclusters is rather similar. The reason for this difference 
between the real and model superclusters is not yet clear; one possible
explanation is that the luminosity-density correlation is not modeled well. 
In Paper I we showed that, on large scales, the luminosity--density 
relation is built into the Millennium Simulation galaxy sample. However, at 
small scales there are large differences of clumpiness between these 
samples; 
this represents the largest difference between the real 
and simulated galaxy populations found in this paper.

\section{Discussion}

\subsection{Shapes and sizes of superclusters}
 
To characterize the shape of an object Sahni et al. (\cite{sah98}),
Sheth et al.  (\cite{sheth03}) and Shandarin et al.  (\cite{sss04})
have investigated the morphology of simulated superclusters and voids 
using the Minkowski functionals and shapefinders.  They studied, among
others, the largest (percolating) supercluster, and showed that
according to shapefinders, this system is filamentary. Sheth et
al. (\cite{sheth03}) plotted the morphology of the largest
(percolating) supercluster in the shapefinder $K_1$-$K_2$ plane for a
limited interval of threshold densities. In this paper the shapefinder
$H_3$ was defined differently from their earlier definition. However,
if we recalculate our shapefinder in the same way as Sheth et al.
(\cite{sheth03}), we get rather similar 
shape curves as their Fig. 18.
 
The shapes and sizes of the observed superclusters were studied by
Basilakos et al. (\cite{bpr01}), Kolokotronis et al. (\cite{kbp02}),
and by Basilakos (\cite{bas03}), and those of LCDM superclusters by
Basilakos et al.  (\cite{bas06}) using Minkowski functionals and
shapefinders for the density field, that was smoothed with a Gaussian
kernel. Kolokotronis et al. (\cite{kbp02}) calculated the shapefinders
$K_{1}$, $K_{2}$, and their ratio $K_{1}/K_{2}$ for Abell
superclusters and showed that about 50\% of all superclusters have the
ratio $K_{1}/K_{2} < 1$ as is typical for filaments. They also showed
that the ratio $K_{1}/K_{2}$ is larger for poor superclusters, which
are typically planar structures (pancakes), and smaller for rich
superclusters, which are more filamentary. This agrees also with our
study which showed that the richest superclusters are multi-branching
filaments. Similarly, Basilakos (\cite{bas03}) showed that at least
70\% of superclusters from 
their SDSS supercluster catalogue are of
filamentary type (the shapefinder $K_{1}/K_{2}<1$). They showed that
also in models filamentary superclusters dominate.

We expand this approach by using the Minkowski functionals and shapefinders to
analyze the full density distribution in superclusters, at all density levels.

In paper II we presented a detailed comparison of sizes and shapes of
superclusters using a number of parameters. First of all, we
calculated the maximum and effective diameters of superclusters (the
maximum diameter is the maximum distance between the grid vertices
belonging to the supercluster, and the effective diameter is the
diameter of a sphere with a volume equal to that of the
supercluster). In order to characterize the compactness of the
supercluster, we used the ratio of these two diameters, the maximum to
the effective. This ratio is the larger, the more empty space is
located in the sphere circumscribed around the supercluster (the more
filamentary the supercluster is). We showed that superclusters are
mainly filamentary objects.  Also we calculated the distance between
the dynamic centers of the superclusters (defined by the position of
the highest density cluster), and the geometric center (defined by the
mean of the maximum coordinate values along the $x,~y,~z-$axes); this
parameter characterizes the asymmetry of superclusters.  Our study
showed that rich superclusters are more filamentary, less compact and
more asymmetrical than poor superclusters.

The Local Supercluster as the closest and best studied supercluster
serves as a supercluster template for poor superclusters, which are
the most numerous superclusters in our catalogue. As seen in the
morphology figures, its morphology is still different from that of
other superclusters in the present selection (see, especially, the
$K_1/K_2$ ratio and its morphological signature in the $K_1$-$K_2$
plane).

\subsection{The peculiar supercluster SCL126}

One of the two richest superclusters from the 2dF survey catalogue is
the supercluster SCL126 in the Northern sky.  In the
$K_1$-$K_2$ shapefinder plane this supercluster is modeled  as a
multi-branching filament.  The $V_{3}$ curve for SCL126 has peaks at
very high mass fractions ($m_f > 0.95$) -- an indication of a high
density compact core. In Einasto et al. (\cite{e03d}) we showed that
the core region of this supercluster contains several Abell clusters,
which are also X-ray clusters. This region has a size of about
10~\Mpc. This supercluster is located almost perpendicularly to the
line-of-sight (Jaaniste et al.  \cite{ja98}, Einasto et al.
\cite{e03d}). In paper RII 
we shall show that the fraction of star-forming
galaxies, especially in the core region, in this supercluster is lower
than in the supercluster SCL9. A possible interpretation of these
findings is that this supercluster started to form earlier than the
supercluster SCL9.

The Minkowski functionals and shapefinders indicate that this
supercluster resembles a rich filament with several branches, and is
less clumpy than other richest superclusters (SCL9, and simulated
superclusters). This interpretation agrees with the description of
this supercluster as a wall in other papers (the Sloan Great Wall,
Vogeley et al. \cite{vogeley04}, Gott et al. \cite{gott05} and
\cite{gott06}, Nichol et al.  \cite{nichol06}). This supercluster
affects the measurements of the correlation function (Croton et
al. \cite{croton04}), and the genus and Minkowski functionals of the
SDSS and 2dF redshift surveys (Park et al.  \cite{park05}; Saar et
al. \cite{saar06}). The "meatball" shift in the measurements of the
topology in the SDSS data is partly due to this supercluster (Gott et
al.  \cite{gott06}). Gott et al. conclude that N-body simulations with
very large volume and more power at large scales are needed to model
such structures more accurately than present simulations.  Similar
conclusions were reached by Einasto et al. (\cite{e06c}).

\section{Conclusions}

We used a catalogue of superclusters of galaxies for the 2dF Galaxy
Redshift Survey and a catalogue of model superclusters from the
Millennium Simulation to study the morphology and internal structure
of the richest superclusters.  Our main conclusions are the following.

\begin{itemize}
  
\item{} The morphology of superclusters can be quantified, using the Minkowski
  functionals and shapefinders.
  
\item{} The fourth Minkowski functional $V_{3}$ describes well the
  clumpiness of superclusters. The value of $V_{3}$ indicates that the
  supercluster SCL126 resembles a multibranching filament, while the
  supercluster SCL9 can be described as a collection of spiders (a
  multispider), consisting of a large number of cores connected by
  relatively thin filaments. Simulated superclusters (especially M2)
  have $V_3$ curves that are somewhat different from those for
  observed superclusters.
  
\item{} We show, using empirical geometrical models, that the
  trajectory traced by the supercluster when we change the mass
  fraction, forms a curve in the $K_1$-$K_2$ plane (the morphological
  signature), which is characteristic to multibranching filaments.
  
\item{} The Minkowski functionals and shapefinders for observed
  superclusters have a much larger scatter than those for simulated
  superclusters.  The Millennium superclusters 
  show very similar 
  morphological scaling relations ($H_1(m_f), H_2(m_f), K_1(m_f),
  K_1/K_2(m_f)$), while these curves vary considerably for the observed
  superclusters.

\item{} The values of the fourth Minkowski Functional $V_{3}$ show that the
  clumpiness of real superclusters, for galaxies of different luminosity, has
  a much larger scatter than the clumpiness of model superclusters. This may
  be an indication that the luminosity-density relation in the models does not
  reflect well the real situation.

\end{itemize}

The present analysis supplements the previous work, adding more
details and using the Minkowski functionals in a novel way.  In
summary, different methods describe together many aspects of the
morphology of superclusters -- their sizes, shapes, volumes,
compactness and clumpiness, giving an overall picture of their
morphology.

\begin{acknowledgements}

  We are pleased to thank the 2dFGRS Team for the publicly available 
  data releases.  We thank T\~onu Viik for helpful suggestions. The 
  present study was supported by the Estonian Science Foundation grants 
  No.  6104 and 7146, and by the Estonian Ministry for Education and 
  Science research project TO 0060058S98. This work has also been 
  supported by the University of Valencia through a visiting 
  professorship for Enn Saar and by the Spanish MCyT project AYA2003-
  08739-C02-01 (including FEDER).  J.E.  thanks Astrophysikalisches 
  Institut Potsdam (using DFG-grant 436 EST 17/4/06),  and the Aspen 
  Center for Physics for hospitality,  where part of this study was 
  performed. 
PH and PN were supported by Planck science in Mets\"ahovi, Academy of Finland.
   In this paper we made use of R, a language for data 
  analysis and graphics (Ihaka \& Gentleman \cite{ig96}).

\end{acknowledgements}

\begin{appendix}
\section{Morphological templates} 
\label{sec:proto} 

While shapefinders have been used in a number of papers, their meaning
needs clarification. In the original paper (Sahni et al. \cite{sah98})
they were illustrated by calculating them for regular geometric bodies
(ellipsoids and tori).  The superclusters we find from observations
are more complex, frequently branching structures, and kernel
estimators give us, as a rule, corrugated density isosurfaces. So, in
order to understand better the shapefinders, we calculate them first
for a well-known household object, a common kitchen table, and then
for several possible supercluster templates.

\begin{figure*}
\centering
\resizebox{0.28\textwidth}{!}{\includegraphics*{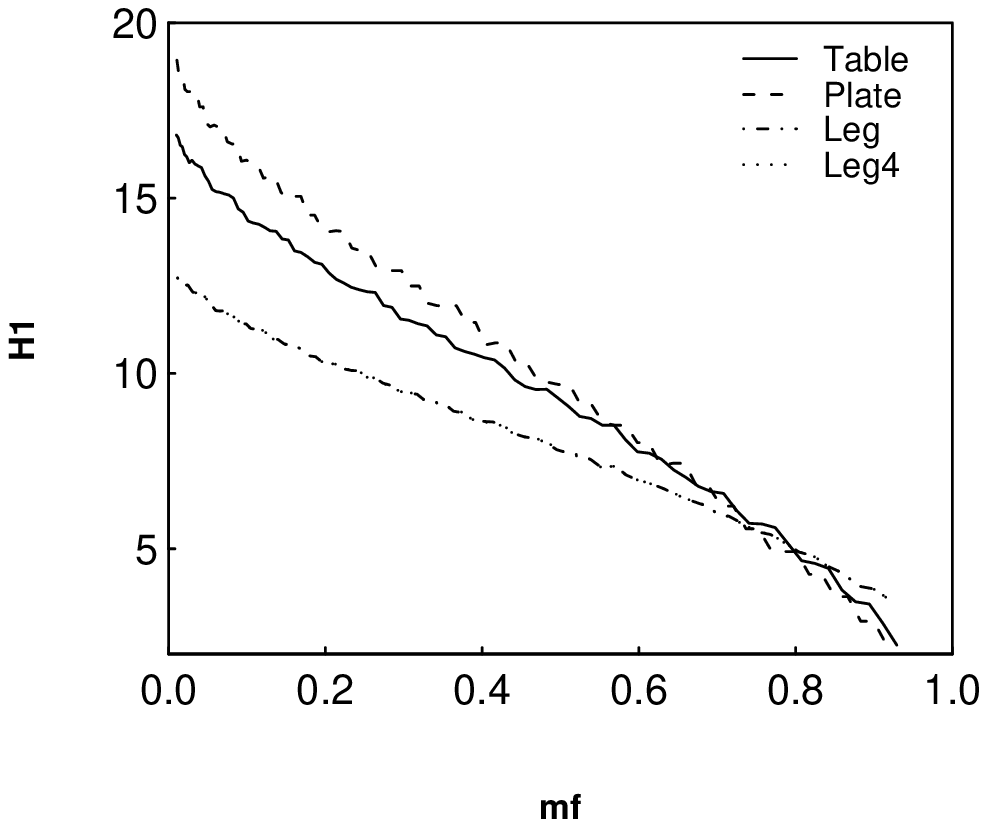}}
\resizebox{0.28\textwidth}{!}{\includegraphics*{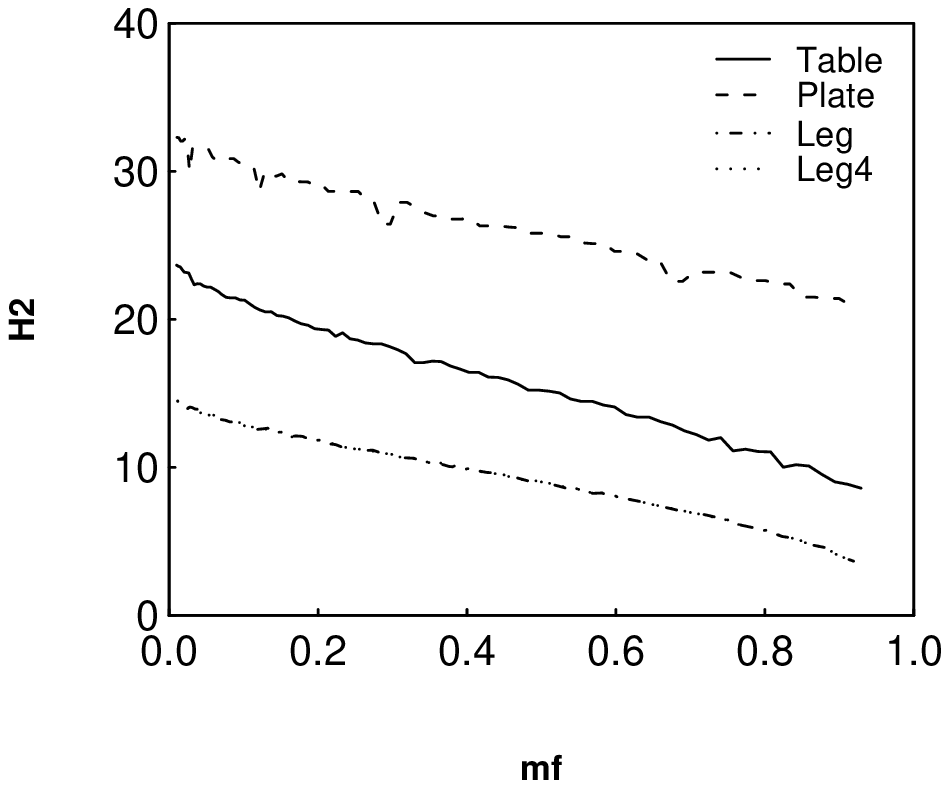}}
\resizebox{0.28\textwidth}{!}{\includegraphics*{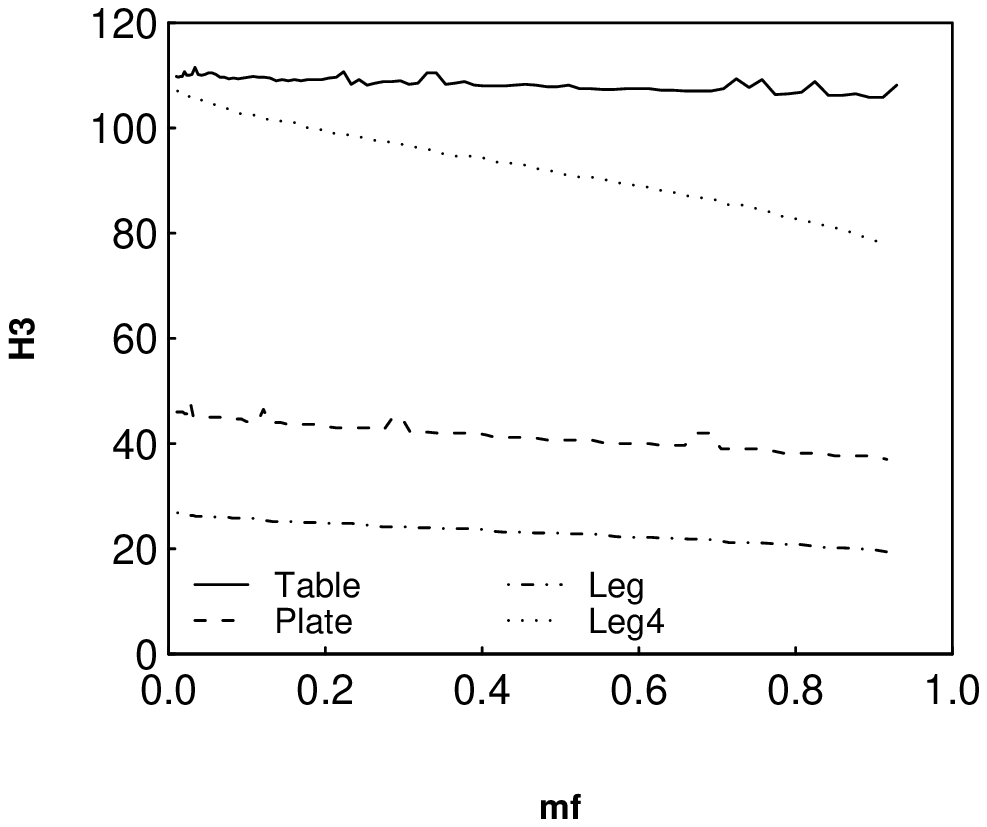}}\\
\resizebox{0.28\textwidth}{!}{\includegraphics*{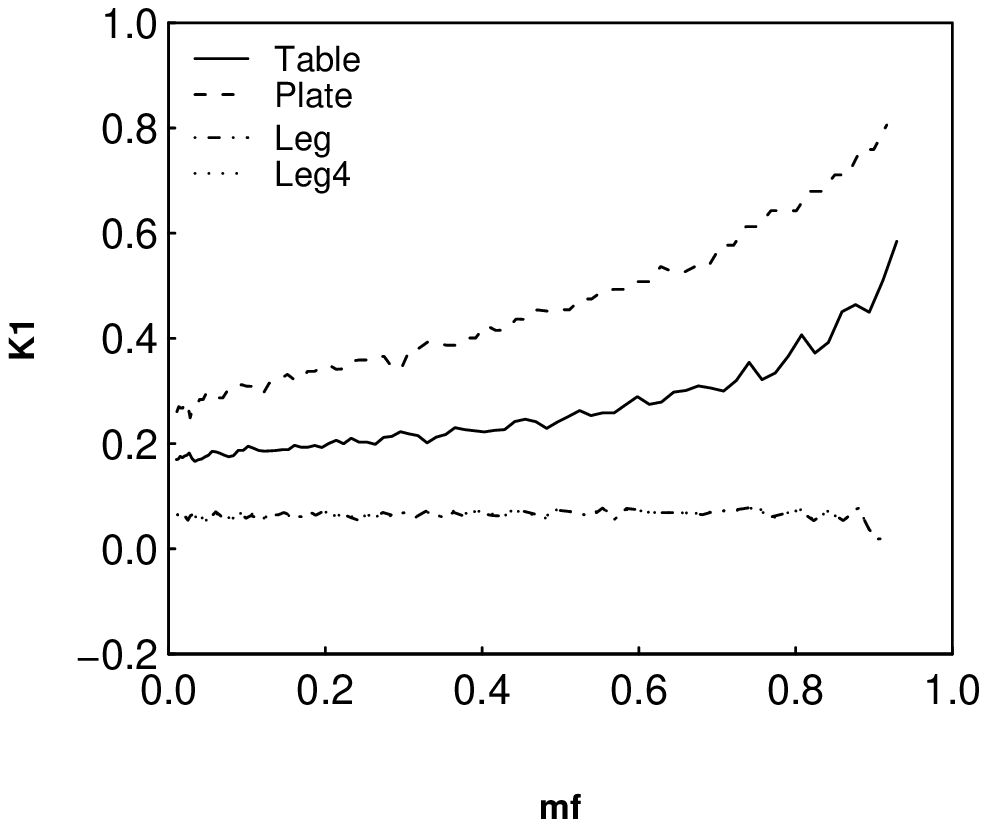}}
\resizebox{0.28\textwidth}{!}{\includegraphics*{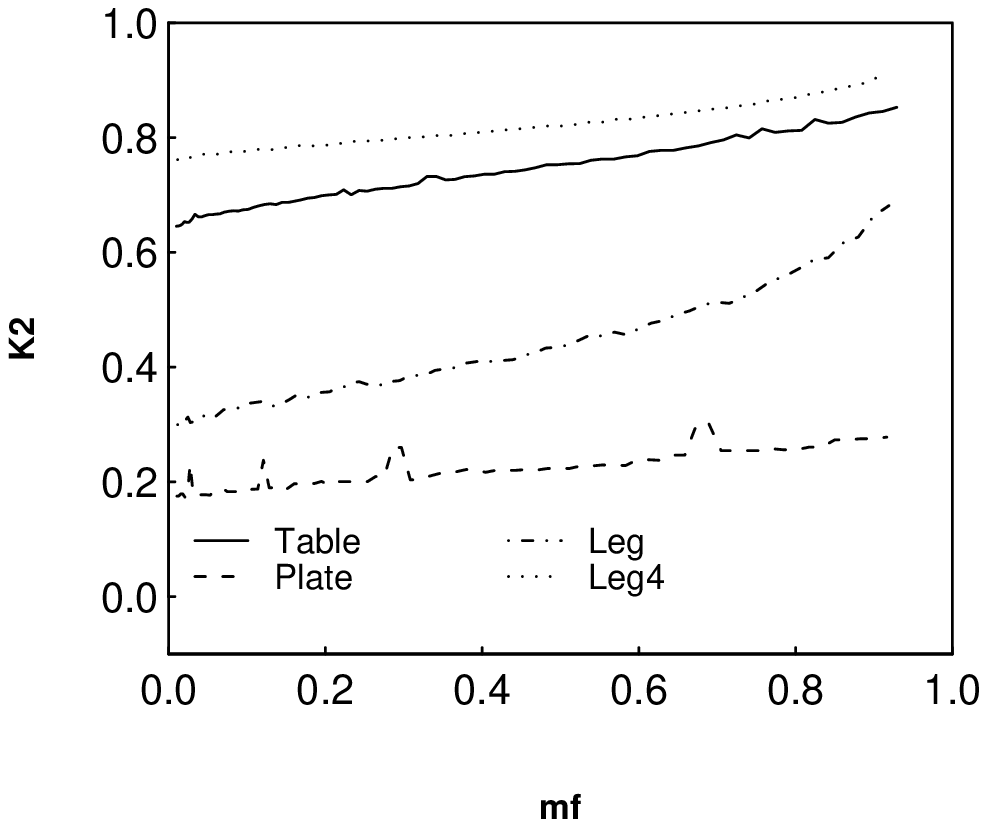}}
\resizebox{0.28\textwidth}{!}{\includegraphics*{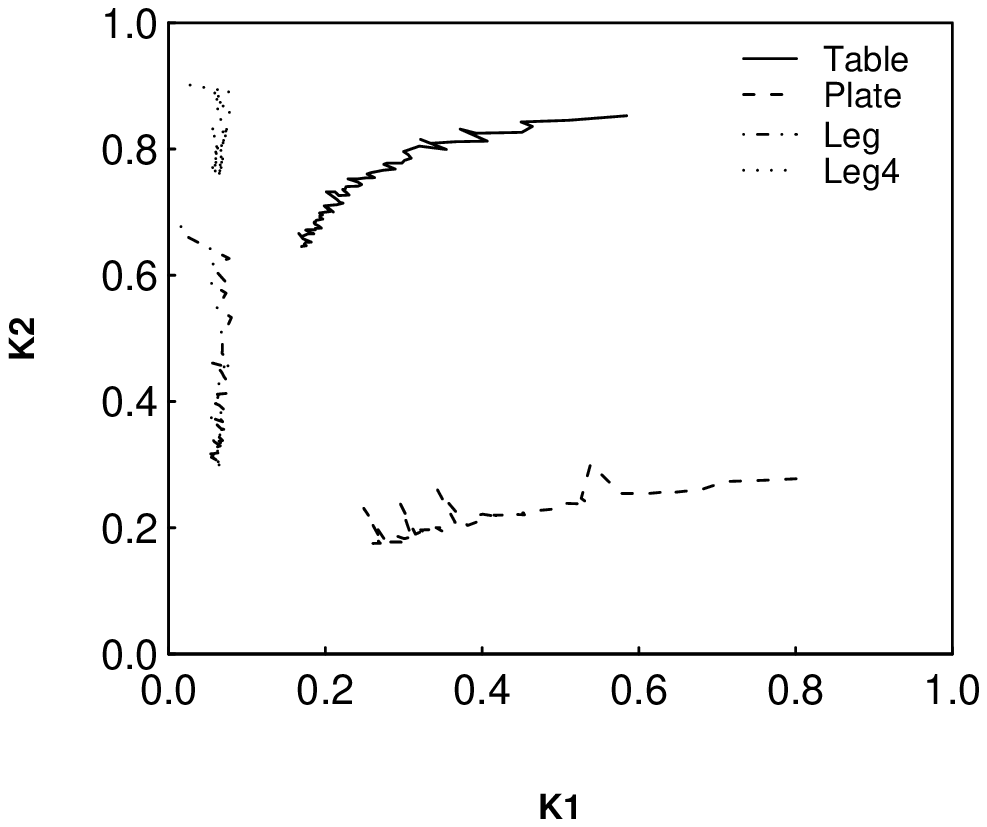}}
\caption{The shapefinders $H_1$ (thickness), $H_2$ (width), $H_3$
(length), upper row, and $K_1$ (planarity), $K_2$ (filamentarity),
and the morphological signature for the table and its details.
\label{fig:table}}
\end{figure*}

We build the table by adding layers of different density to a
skeleton, consisting of a plate of the size of $40\times 100$ (grid
units), and of four legs (rods) of the length of $60$, which join the
plate perpendicularly at its corners.  We use the parabolic density
law:
\[
\varrho(r)=1-r^2/R^2,\quad r\leq R; \varrho(r)=0,\quad  \mathrm{otherwise},
\]
where $r$ is the nearest distance from a point to the skeleton,
and we have taken the limiting distance $R=10$ grid units. This is a rather
thick table with thick legs; the reason for that is the following.
First, as we use Crofton formulae to calculate the Minkowski functionals,
we have to ensure that our table occupies a generic position in space
(the specific weights we use are obtained by assuming statistical isotropy
of the isodensity surfaces, see Schmalzing and Buchert (\cite{jens97})).
We do this by aligning the normal to the plate in the direction
${\mathbf n}=(1/\sqrt{3},1/\sqrt{3},1/\sqrt{3})$. As the plate is
inclined with the respect to the coordinate grid, the isosurfaces are
inevitably slightly jagged, and large $R$ makes them smoother.
When following the density isosurfaces inwards, towards larger $m_f$,
we find another effect -- because of the inclination of the skeleton
the high density 
regions on the grid break up into a large number of isolated clumps
around grid points. So we shall present the shapefinders only for
mass fractions $m_f\leq 0.93$ (up to the break-up). The shapefinders are 
given in Fig.~\ref{fig:table}, for the full table, for the plate, for
a single leg and for four legs together.

Let us start with the first shapefinder, the thickness $H_1$. We know
by construction that the true thickness is $2R=20$ at the outer edges
of the table, and reaches 0 inside. The Figure shows that $H_1$ almost
finds the true thickness for the plate, is slightly lower for the
leg(s), starting from 13, and runs in between those two relations for
the full table.

The second shapefinder, $H_2$, should give the width of the table. 
For the plate it should be $40+2R=60$ at low densities and 40 at high
densities. As we see, it runs from about 33 to 22, more close to the
half-width of the plate; this is logical, considering that the shapefinders
are normalized requesting that the 'width' of a sphere should equal its
radius, not diameter. But a similar normalization is used for the
thickness above, which, however, finds the full thickness.
The width of the leg(s) is equal to their thickness, $H_1=H_2$ -- the legs
are cylinders. And the 'width' (half-width, to be more precise) of the table, 
the combination of the above details, is closer to the widths of the legs;
these four appendices dominate the width shapefinder for the full table.

The third shapefinder $H_3$ should give the length of the objects. It
starts from about 47 for the plate, and from 28 for a leg. Here,
finally, the shapefinder for four legs is four times that for a single
leg, as it should be. But these values show that $H_3$ gives really
the half-length of the object, because of the normalization,
again. And the half-length of the full table is larger than any of its
details, but does not equal their sum. Moreover, while the lengths of
the details approach those of the skeleton, the length of the full
table stays, surprisingly, the constant.

The 'second-order' shapefinder, the planarity $K_1$, is small (about
0.06) for the leg(s), runs from 0.27 (thick plate) to 0.8 (fat
skeleton) for the plate, and from 0.18 to 0.6 for the full
table. These numbers are as expected, only the maximum planarity for
the plate seems to be too small; but, as explained above, this might
be simply a non-aligned grid effect.

The filamentarity $K_2$ is smallest for the plate, ranging from 0.17
to 0.27.  It is not too large for a single leg (from 0.3 to 0.7, the
legs are thick), but ranges from 0.76 to 0.92 for the collection of
four legs. The filamentarity of the full table is dominated by that of
the legs; it ranges from 0.64 to 0.85; a kitchen table is basically a
filamentary object.

The planarity/filamentarity ratio $K_1/K_2$ that is frequently used,
runs from 0.2 to about 0.1 for a single leg, and is lower than 0.07
for the collection of four legs.  It is high (larger than 1.5) for the
plate, but pretty low (between 0.5 and 0.7) for the full table;
another indication that legs define the morphology of a table.  This
curve is pretty jaggy, and the reason for that is clear -- the 'first
order' shapefinders $H_i$ are ratios of 
Minkowski functionals, the 'second order' shapefinders $K_i$ are ratios of
these ratios, and the ratio $K_1/K_2$ become extremely noisy.

We can also look at the morphological signatures (the trajectories in 
the $K_1$--$K_2$ plane, parameterized by $m_f$) of the table and its 
details (Fig.~\ref{fig:table}, lower right panel). We limit the mass 
fraction range for this figure also from below, using only 
$m_f\in[0.01,0.94]$, as done in the main paper (see also appendix B). 
This choice eliminates grid effects at low densities.

Ignoring the jaggy nature of the curves (these can be eliminated, in
principle, by constructing direct estimates for the shapefinders $K_1$
and $K_2$) we see that the details and the table have clear and
distinct signatures.  The signature for the plate lies in the planar
region, that for the leg(s) in the region describing filaments (and
four legs together are clearly more filamentary than a single
leg). The full table is both planar $K_1\in[0.2,0.7]$ and filamentary
$K_2\in[0.65,0.9]$, 
but clearly rather filamentary than planar ($K_2 > K_1$).

\begin{figure*}
\centering
\resizebox{0.28\textwidth}{!}{\includegraphics*{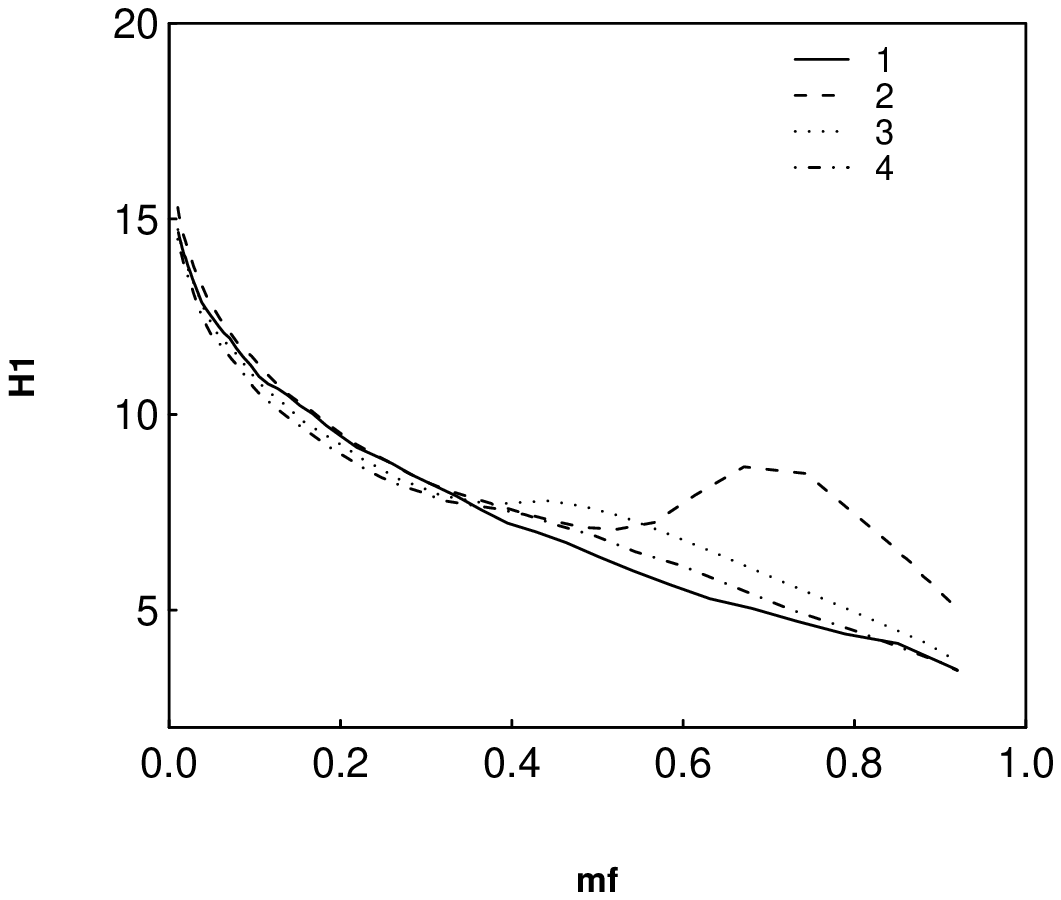}}
\resizebox{0.28\textwidth}{!}{\includegraphics*{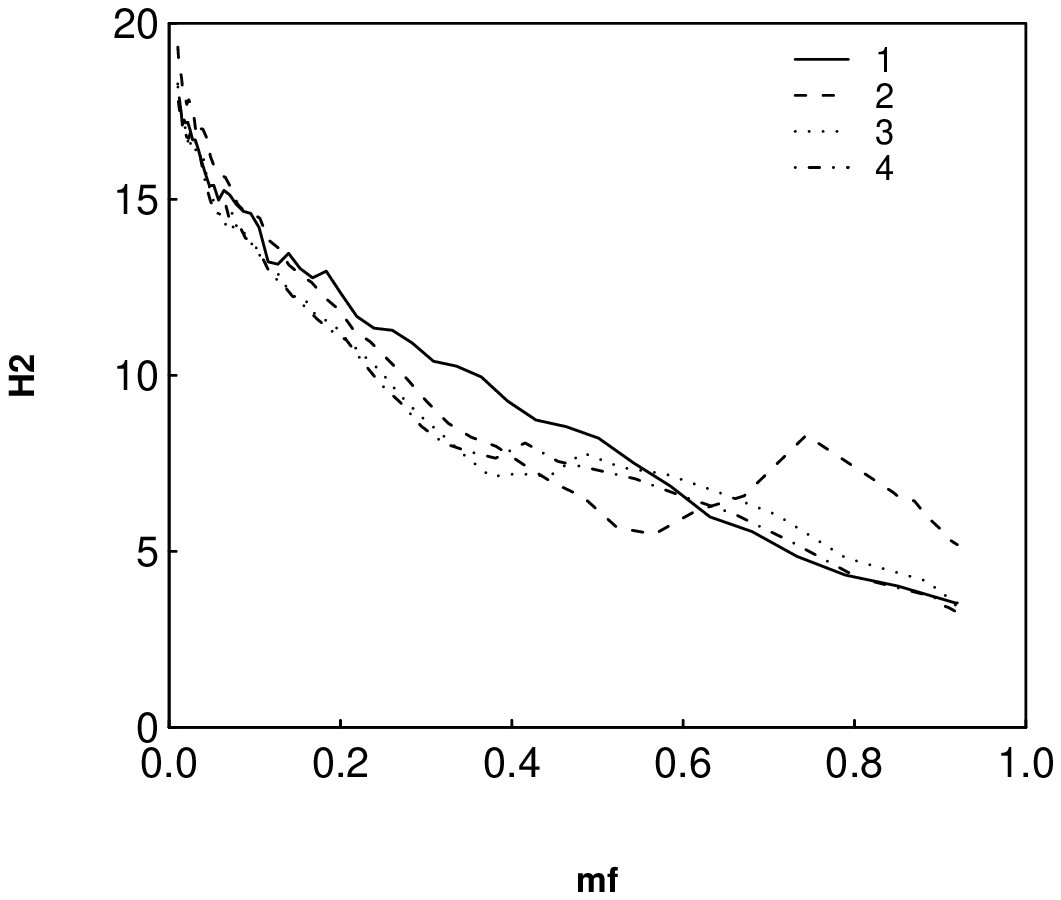}}
\resizebox{0.28\textwidth}{!}{\includegraphics*{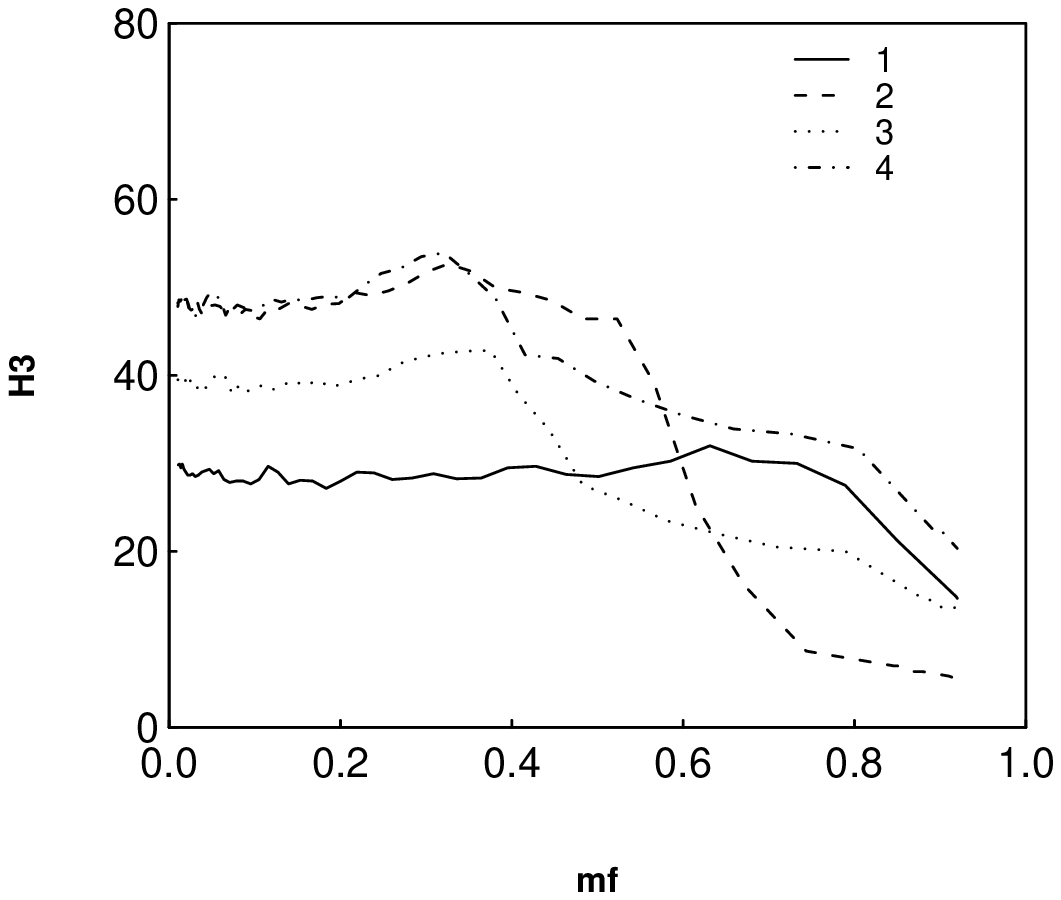}}\\
\resizebox{0.28\textwidth}{!}{\includegraphics*{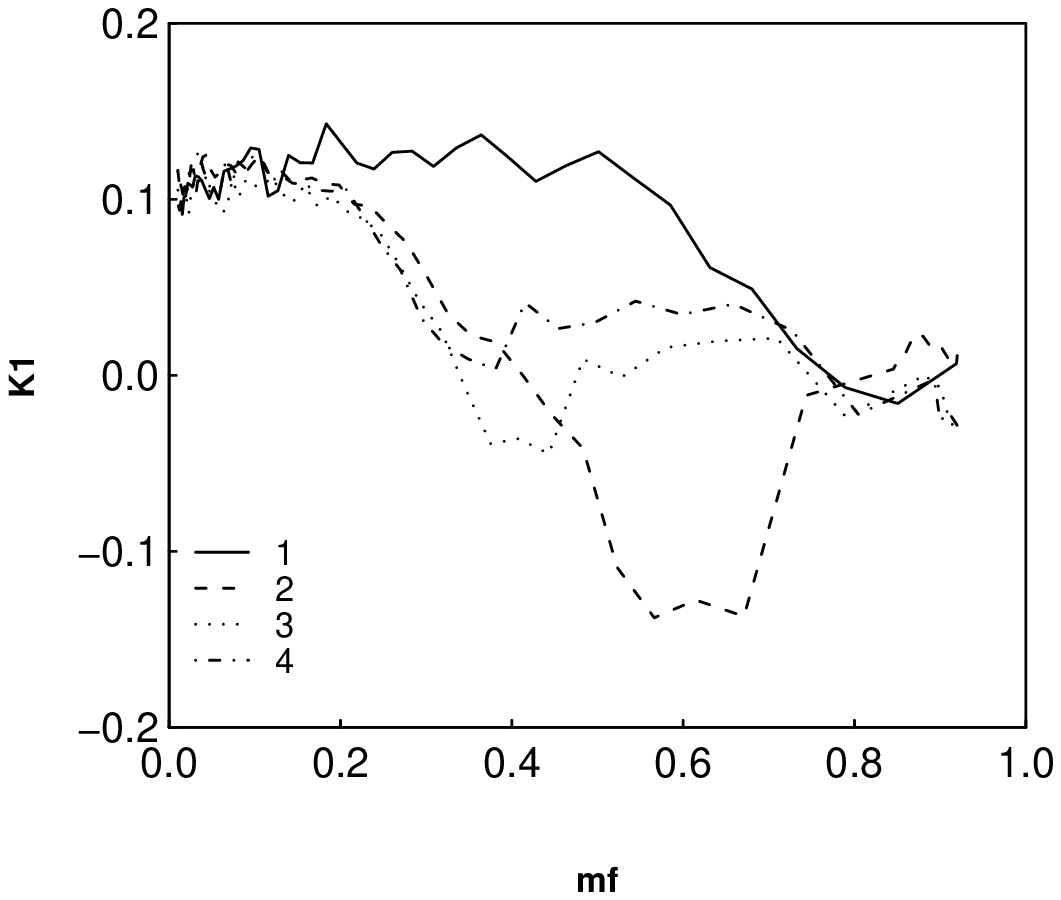}}
\resizebox{0.28\textwidth}{!}{\includegraphics*{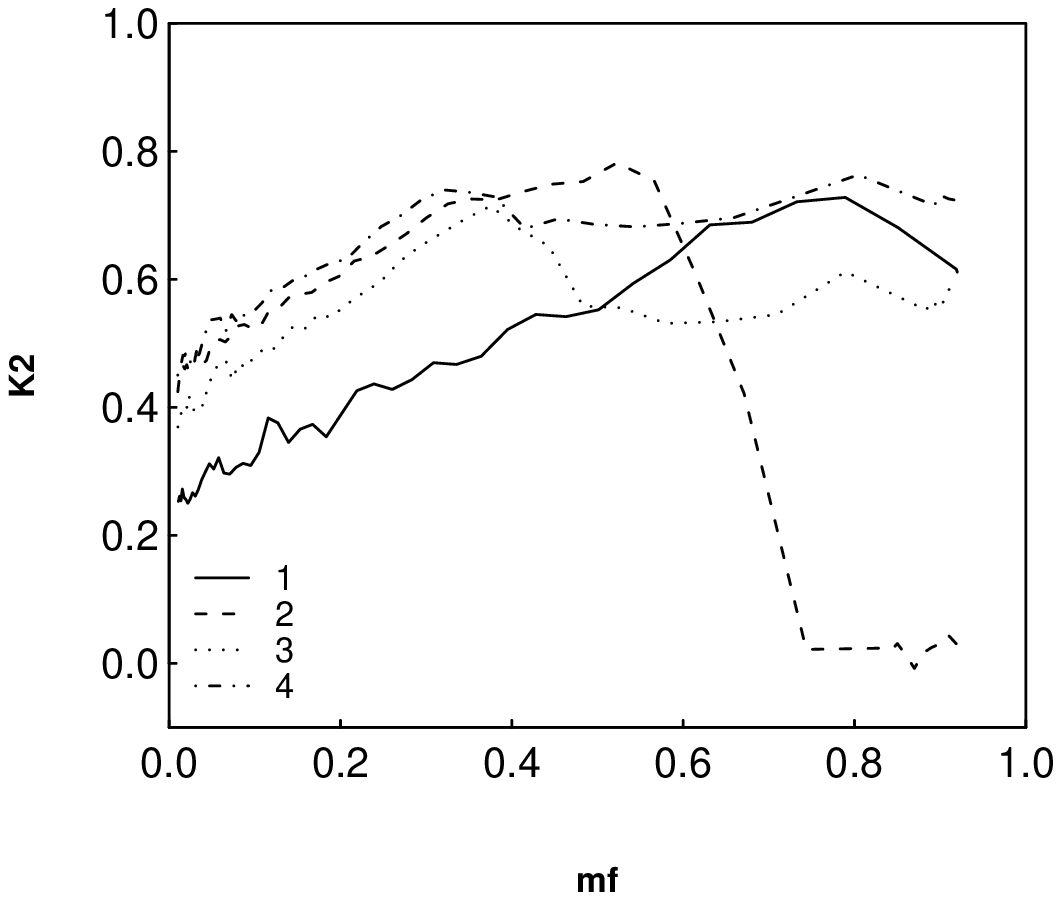}}
\resizebox{0.28\textwidth}{!}{\includegraphics*{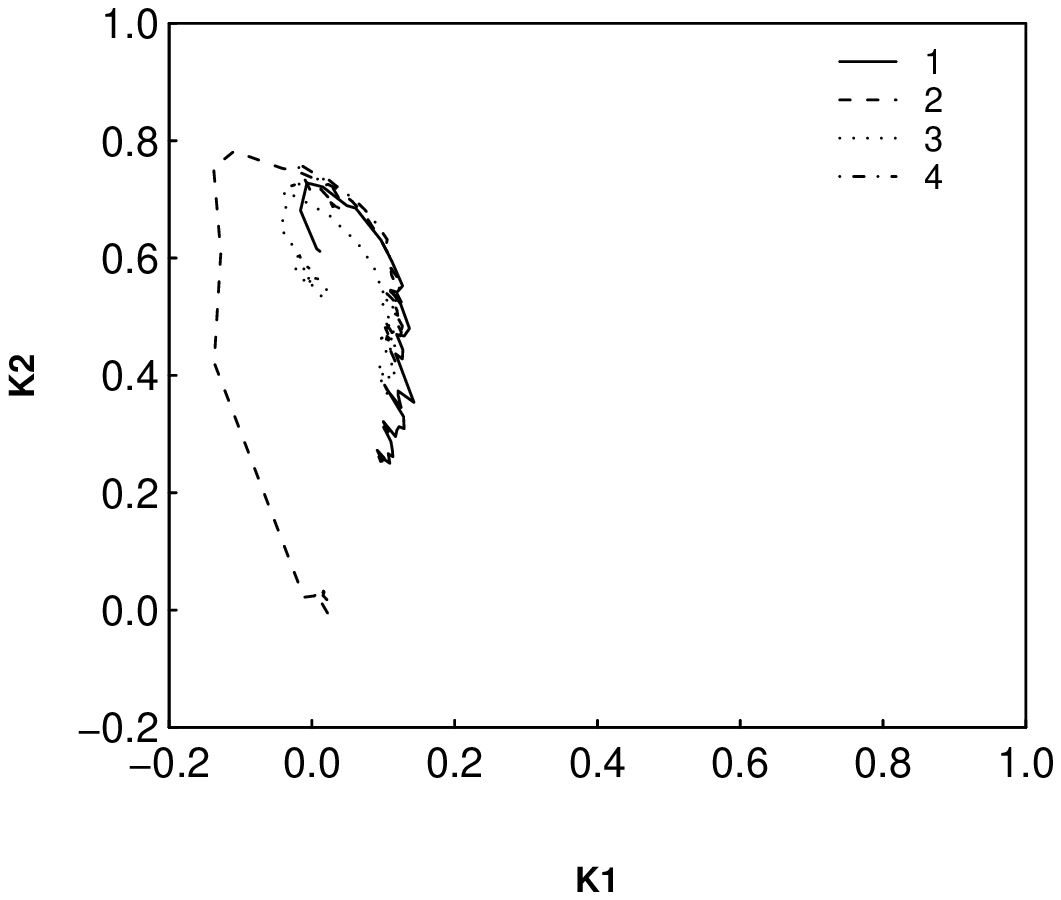}}
\caption{The shapefinders $H_1$ (thickness), $H_2$ (width), $H_3$
(length), upper row, and $K_1$ (planarity), $K_2$ (filamentarity),
and the morphological signature (lower row) for different templates.
The labels are: 1 -- filament, 2 -- spider, 3 -- Jacob's
staff, and 4 -- Cross of Lorraine (Karelian reindeer). 
\label{fig:proto}}
\end{figure*}

The kitchen table was an example of a smooth density distribution. The
density of our superclusters is obtained by summing together almost
isotropic kernels around galaxy locations. This results in a much more
patchy density distribution; on the other hand, it automatically
guarantees statistical isotropy of the density isolevels. In order to
better understand the shapefinders of the observed superclusters, we
calculate them for a number of simple templates that resemble the
superclusters. These are: 1) a single long filament, 2) a spider (two
long filaments, crossing, with a small cube at the centre), 3) Jacob's
staff or ballestina (an astronomical instrument from middle ages for
measuring angular distances; a long staff with a perpendicular smaller
staff), and 4) the Cross of Lorraine (similar to the Jacob's staff, but
with two perpendicular staves, with a spherical blob at the centre).
The last configuration resembles also ancient petroglyphs from
Karelia, depicting thunder and reindeer.  The dimensions are: 100
(grid units) for the filament (and staff) length, 40 for the small
staff length, 10 for the radius of the blob, and the effective width
of the density kernel is 8 (total width 16).  The corresponding
shapefinders are shown in Fig.~\ref{fig:proto}.

As we see, the shapefinders are smoother, and clearly distinct for
different morphologies. The complex (corrugated) nature of the
isodensity surfaces causes local minima for the lengths $H_i$ at
certain mass fractions (in the thickness and in the width of the
spider). Also, the length shapefinder $H_3$ almost does not change for
a wide initial interval of mass fractions, while the isodensity
surfaces contract towards the skeletons of our templates. As this
length is defined both by the isodensity surface area and its
curvature, 
a simple behaviour cannot be always expected.

A new feature here is that the 'second order' shapefinder $K_1$
acquires negative values for specific density isolevels. This is
possible, as the inequalities $H_1\leq H_2\leq H_3$ that are usually
assumed, are valid only for convex bodies. The isolevels for
kernel-estimated densities are usually corrugated, and the real limits
for $K_i$ are: $K_i\in[-1,1] $.  This may forbid the use of the usual
morphological index $K_1/K_2$; it is better to use the 'morphological
signature' $K_2$ versus $K_1$ instead.

Looking at Fig.~\ref{fig:proto} we see that there are two morphologies
that resemble those of the observed superclusters. The 'spider'
imitates poor Virgo-type superclusters, although our model spider has
a large value $K_2$, rich superclusters (as SCL126, see
Fig.~\ref{fig:mf4k1k2}) are best represented by the Cross of
Lorraine, a multibranching filament. Adding a central body to that
system improves the similarity yet more.

We do not attempt to fit supercluster models to observations in this
paper; this will need a well-parameterized basic model. However, we
see that a good starting point for such a project would be the two
models listed above.

\section{Kernel densities}
\label{sec:kernel}

When studying the morphology of superclusters of galaxies, a necessary
step is to convert the spatial positions of galaxies into spatial densities.
The standard approach is to use kernel densities (see, e.g., Silverman 
\cite{silverman}):
\[
\varrho(\mathbf{x})=\sum_i K(\mathbf{x-x}_i;h) m_i,
\]
where the sum is over all galaxies, $\mathbf{x}_i$ are the coordinates
of the $i$-th galaxy, and $m_i$ is its mass (or luminosity, if we are
estimating luminosity densities; 
for number densities we set 
$m_i\equiv1$).  The function $K(\mathbf{x};h)$ is the kernel of the
width $h$, and a suitable choice of the kernel determines the quality
of the density estimate. If the kernels widths depend either on the
coordinate $\mathbf{x}$ or on the position of the $i$-th galaxy, the
densities are adaptive. We are using constant width kernels in this
paper, defining superclusters as density enhancements of a common
scale (the typical density scale of about 8\hmpc).  Kernels have to be
normalized and symmetrical:
\[
\int K(\mathbf{x};h)dV=1,\qquad \int \mathbf{x}K(\mathbf{x};h)dV=0.
\]

Statisticians classify kernels by their MSE (minimal standard error); the
best kernel is the Epanechnikov kernel $K_E$:
\[
K_E(\mathbf{x};h)=\frac{15}{8\pi h^3}(1-\mathbf{x}^2/h^2)\;,\quad \mathbf{x}^2\leq h^2,\; 
\quad 0\quad \mathrm{otherwise}\;.  
\]
(all the kernel expressions in this appendix are for 3-D kernels).

In cosmology, the Gaussian kernel $K_G$ has been the most popular:
\begin{equation}
\label{Gauss}
K_G(\mathbf{x};h)=\frac1{(2\pi)^{3/2}h^3}\exp\left(-\mathbf{x}^2/2h^2\right)\;.
\end{equation}

\begin{figure}
\centering
\resizebox{0.45\textwidth}{!}{\includegraphics*{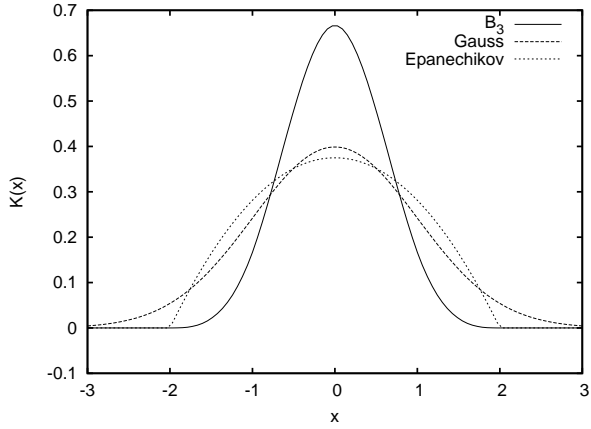}}
\caption{Three popular kernels for density estimation -- Epanechnikov, Gaussian,
and the $B_3$ box spline. All the profiles have approximately the same
extent (the Gaussian profile width $h=1$, the $B_3$ profile is given for $N=1$,
and for the Epanechnikov profile $h=2$. All profiles are normalized
for the 1-D case. 
\label{fig:appb3} }
\end{figure}

For the usual case, when densities are calculated for a spatial grid,
good kernels are generated by box splines $B_J$ (these are usually
used in $N$-body mass assignment).  Box splines have compact support
(they are local), and they are interpolating on a grid:
\[
\sum_i B_J(x-i)=1,
\]
for any $x$, and a small number of indices that give non-zero values for
$B_J(x)$.  In this paper we restrict us to
the popular $B_3$ splines:
\[
B_3(x)=\frac1{12}\left[|x-2|^3-4|x-1|^3+6|x|^3-4|x+1|^3+|x+2|^3\right]
\]
(this function is different from zero only in the interval $x\in[-2,2]$,
see Fig~\ref{fig:appb3}).
We define the (one-dimensional) $B_3$ box spline kernel 
of width $h=N$ as
\[
K_B^{(1)}(x;N)=B_3(x/N)/N.
\]
This kernel preserves the interpolation property (mass conservation)
for all kernel widths that are integer multiples of the grid step, $h=N$.
The 3-D $K_{B}^{(3)}$ box spline kernel we use is given by the direct 
product of three one-dimensional kernels:
\[
K_B(\mathbf{x};N)\equiv K_B^{(3)}(\mathbf{x};N)=K_B^{(1)})(x;N)K_B^{(1)}(y;N)
        K_B^{(1)}(z;N),
\]
where $\mathbf{x}\equiv\{x,y,z\}$.

These kernels encompass all the good and bad kernel properties (good and
bad for our application).
\begin{itemize}

\item First, the Epanechnikov and the $B_3$ kernels are both compact, while
the Gaussian kernel is infinite and has to be cut off at a fixed radius.
This introduces an extra parameter and, what is more important, generates
small-amplitude density jumps that may distort the Minkowski functionals
(these are extremely sensitive to small-amplitude density details).

\item Second, the $B_3$ kernel conserves density (it 
is exactly normalized on a
grid). For continuous kernels normalization on a grid is equivalent to
simple numerical integration of the profile; the (truncated) Gaussian
kernel demands less grid points for that than the Epanechnikov kernel.
Many grid points inside the kernel profile means that kernel widths
must be much larger than the grid step.
Bad normalization leads to distortion of the final densities; how well the
contribution of a galaxy to the final density is accounted for depends on the
location of the galaxy with respect to the grid.

\item Third, an important kernel property is its isotropy. The formulae that we
use to estimate the Minkowski functionals assume average isotropy of the
density isosurfaces (their isotropic orientation). Both the Epanechnikov
and Gaussian kernels are exactly isotropic, the 3-D $B_3$ kernel is not.
But this kernel is approximately isotropic to a high degree. 
\end{itemize}

\begin{figure}
\centering
\resizebox{0.45\textwidth}{!}{\includegraphics*{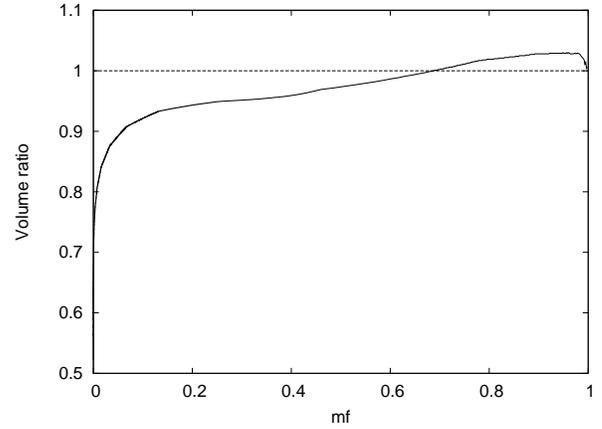}}
\caption{The ratio of the volume inside an isolevel to the volume of the
sphere that cuts coordinate axes at the same point as the isolevel (an
inscribed sphere for larger radii, an encompassing sphere for smaller radii).
The isolevels are labeled by the 'mass fraction', the kernel mass outside
the isolevel. For an isotropic kernel this ratio should be 1.
\label{fig:appiso} }
\end{figure}

A simple characteristic of isotropy of the kernel isolevels would be
the ratio of its surface to the surface of the sphere that cuts
coordinate axes at the same point as the isolevel. Alas, surfaces are
not easy to calculate, and we replace this ratio by the ratio of the
volumes encompassed by these two surfaces; volumes are easy to
estimate by Monte-Carlo integration. These ratios are close (e.g., for
a cube and the inscribed sphere the volume ratios coincide with the
surface ratios).  The dependence of this ratio (the sphere volume to
the excursion set volume) as a function of the mass fraction
(integrated kernel mass outside the isolevel) is shown in
Fig.~\ref{fig:appiso}.

As we see, this ratio is close to unity for almost all mass fractions;
considerable deviations can be seen only for small mass fractions.
The deviation from isotropy is less than 10\% for $m_f\geq 0.058$.
When displaying our results on the morphological descriptors, we used a
mass fraction limit for the full supercluster; this is smaller than the
value cited above by the ratio of the number of galaxies defining the 
supercluster limits to the total number of galaxies in the supercluster.
To be on a safe side, we knowingly underestimated this ratio and we
chose the limit $m_f\geq 0.01$ for the full supercluster.

\end{appendix}
\end{document}